\pdfoutput=1
\documentclass[12pt]{article}
\usepackage{latexsym} 
\usepackage{verbatim}
\usepackage{tikz}
\usepackage{color}
\usepackage{graphicx,amsfonts,amsmath,amssymb,amscd,amstext,mathrsfs}
\usepackage{graphicx} 
\usepackage{bbm}
\usepackage{dsfont}

\usepackage{xcolor}
\definecolor{DarkBlue}{rgb}{0.0, 0.0, 0.55}
\usepackage[colorlinks=true, citecolor=DarkBlue, linkcolor=DarkBlue, allcolors=DarkBlue, linktoc=all]{hyperref}   

\usepackage{empheq}
\usepackage{hyperref}
\usepackage{lipsum}

\usepackage[title,titletoc]{appendix}
\usepackage{tocloft}

\def\be{\begin{eqnarray}}
\def\ee{\end{eqnarray}}
\def \bea {\begin{equation}}
\def \eea {\end{equation}}
\def\frac#1#2{{#1\over #2}}
\def\nn{\nonumber}

\DeclareFontShape{OT1}{cmr}{mx}{n}{<->cmr10}{}

\def \z{\bar z}

\usepackage{cite}

\begin{document}

\begin{titlepage}

\begin{flushright} 
\end{flushright}

\begin{center} 
\vspace{1cm}  

{\fontsize{16pt}{0pt}{\bf 
 Resummation of Multi-Stress Tensors\\ 
\vspace{0.35cm} 
in Higher Dimensions
}}
\vspace{1.7cm}  
\\
 {\fontsize{10pt}{10pt}{Kuo-Wei Huang 
}}   
\\ 
\vspace{0.4cm} 
{\fontsize{11pt}{0pt}{\it Mathematical Sciences, University of Southampton, \\
Highfield, Southampton SO17 1BJ, UK}}\\
\end{center}
\vspace{0.6cm} 

\begin{center} 
{\bf Abstract}
\end{center} 
\vspace{-0.05cm} 
{\noindent  In the context of holographic conformal field theories (CFTs), a system of linear partial differential equations was recently proposed to be the higher-dimensional analog of the null-state equations in $d=2$ CFTs at large central charge. 
Solving these equations in a near-lightcone expansion yields solutions that match the minimal-twist multi-stress tensor contributions to a heavy-light four-point correlator (or a thermal two-point correlator) computed using holography, the conformal bootstrap, and other methods.
This paper explores the exact solutions to these equations.  
We begin by observing that, in an expansion in terms of the ratio between the heavy operator's dimension and the central charge, the $d=2$ correlator involving the level-two degenerate scalars at each order can be represented as a Bessel function; the resummation yields the Virasoro vacuum block. 
We next observe a relation between the $d=2$ correlator and the $d=4$ near-lightcone correlator involving light scalars with the same conformal dimension. 
The resummed $d=4$ correlator takes a simple form in the complex frequency domain.  
Unlike the Virasoro vacuum block, the resummation in $d=4$ leads to essential singularities. 
Similar expressions are also obtained when the light scalar's dimension takes other finite values. These CFT results correspond to a holographic computation with a spherical black hole. 
In addition, using the differential equations, we demonstrate that the correlators can be reconstructed via certain modes. In $d=2$, these modes are related to the Virasoro algebra.  
}

\end{titlepage}

\addtolength{\parskip}{1.2 ex}
\jot=0.6 ex 

\hypersetup{linkcolor=black}
\tableofcontents
\newpage

\section{Introduction} 

The aim of the present work is to obtain some analytic, exact structures relevant to a four-point scalar correlator in a class of higher-dimensional Conformal Field Theories (CFTs) that have an Einstein gravity dual description. Rather than performing a gravity computation, our approach here is to solve a set of linear partial differential equations.

\vspace{-0.2cm}

\subsection{Motivations and assumptions} 

A motivation behind this work is the desire to develop field-theoretic approaches to better organise the computational complexity of strongly-coupled quantum field theories (QFTs). From the renormalisation group perspective \cite{WILSON197475}, defining a QFT is equivalent to understanding continuous phase transitions described by CFTs. Since the 1984 landmark paper by Belavin, Polyakov, and Zamolodchikov (BPZ)  \cite{BELAVIN1984333}, which introduced a powerful analytic framework for solving a class of $d=2$ systems, $i.e.$, minimal models, by leveraging the Virasoro symmetry and the conformal bootstrap, the CFT formulation has been extended and further elaborated over the past four decades, yielding a vast array of remarkable results in both two-dimensional and higher-dimensional theories. Recent advances in the conformal bootstrap are reviewed in, $e.g.$, \cite{Poland:2018epd, Bissi:2022mrs, Hartman:2022zik, Rychkov:2023wsd}. 

An analytic approach to $d>2$ CFTs is the lightcone bootstrap, which is based on the universal observation that every CFT has a large-spin expansion accessible via the near-lightcone limit of the crossing equations \cite{Komargodski:2012ek, Fitzpatrick:2012yx}; see also \cite{Alday:2007mf, Pal:2022vqc}. In the near-lightcone regime, the $t$-channel operator product expansion (OPE) is controlled by the minimal-twist operators, where the twist of an operator is the difference between its conformal dimension and its spin. In general, it is challenging to explicitly sum over contributions from all relevant minimal-twist exchanged operators. In this work, we would like to make progress in the context of holographic CFTs.   According to \cite{Heemskerk:2009pn}, these CFTs require a large central charge $C_T\to \infty$, and a large gap to the lightest spin-$\ell >2$ single-trace primary. Related discussions can be found in, $e.g.$, \cite{Fitzpatrick:2010zm, Camanho:2014apa, Hartman:2015lfa, Afkhami-Jeddi:2016ntf, Belin:2019mnx, Kologlu:2019bco, Caron-Huot:2021enk}. In this note, we focus on the contributions from minimal-twist multi-stress tensors, $i.e.$, the stress tensor and its infinite number of composites, to a four-point correlator involving two heavy scalars with a large dimension $\Delta_H \to \infty$, and two light probe scalars with certain restricted, finite dimensions $\Delta_L=\Delta$. 

Another motivation arises from the quest to understand the nature of black holes and quantum gravity via the holographic framework \cite{Maldacena:1997re, Gubser:1998bc, Witten:1998qj} in a higher-dimensional setting. A substantial amount of work has been devoted to the lower-dimensional cases. 
Of particular relevance to the motivation of this note is the exact expression of the $d=2$ Virasoro vacuum block at large central charge, which presents the resummation of all minimal-twist multi-stress tensors in $d=2$ \cite{Fitzpatrick:2014vua, Fitzpatrick:2015zha} and captures many important aspects of two-dimensional CFTs and three-dimensional quantum gravity, $e.g.$, \cite{zamolodchikov1984conformal, zamolodchikov2, Harlow:2011ny, Hartman:2013mia, Litvinov:2013sxa, Asplund:2014coa, Roberts:2014ifa,  Hijano:2015rla, Perlmutter:2015iya, Fitzpatrick:2015qma, Hijano:2015qja, Fitzpatrick:2015foa, Beccaria:2015shq, Fitzpatrick:2015dlt, Alkalaev:2015fbw,  Fitzpatrick:2016thx, Banerjee:2016qca,  Anous:2016kss,   Fitzpatrick:2016ive, Chen:2016cms, Chen:2016dfb,  Maloney:2016kee,  Fitzpatrick:2016mtp,  Chen:2017yze, Cho:2017oxl, Cho:2017fzo,  Galliani:2017jlg, Bombini:2017sge,  Kusuki:2018nms,  Bombini:2018jrg, Cotler:2018zff,  Kusuki:2018wpa,  Collier:2018exn, Kulaxizi:2018dxo, Chang:2018nzm,  Giusto:2018ovt, Anous:2019yku, Kusuki:2019gjs,  Besken:2019bsu,  Haehl:2019eae,  Besken:2019jyw, Collier:2019weq, Alkalaev:2020kxz, Anous:2020vtw,  Das:2020fhs,  Nguyen:2021jja, Karlsson:2021mgg, Nguyen:2022xsw, Benjamin:2023uib, Eberhardt:2023mrq, Grabovsky:2024jwf}.  
In particular, the exchanges of multi-stress tensors in $d=2$ are dual to the exchanges of multi-gravitons in AdS$_3$, describing how light probe fields interact with a heavy object such as a BTZ black hole \cite{Banados:1992wn}. Higher-dimensional CFTs are less constrained, and the dual gravity theories involve propagating gravitational waves, making the higher-dimensional systems more complex. 

In \cite{Fitzpatrick:2019zqz}, a computational scheme was developed to analytically solve the bulk scalar field equation in higher-dimensional AdS gravity with a black-hole background, enabling one to compute multi-stress tensor contributions to the thermal two-point scalar correlator (in position space). The thermal correlator is interpreted as a heavy-light four-point correlator, where two heavy scalar operators create a thermal background. This scheme relies on a near-boundary expansion and a choice of variables, allowing one to also incorporate higher-derivative corrections to the gravitational action. In the case of AdS$_3$, this scheme reproduces the $d=2$ Virasoro vacuum block at large central charge.\footnote{The scheme does not determine the double-trace contributions made out of two light probe scalars, as these contributions depend on an interior boundary condition.} Some extensions can be found in \cite{Li:2019tpf, Fitzpatrick:2019efk, Fitzpatrick:2020yjb, Karlsson:2022osn, Huang:2022vet, Esper:2023jeq, Ceplak:2024bja}. Using the OPE coefficients computed via holography with a spherical black hole, a simple closed-form expression that represents summing over $d=4$ double-stress tensor exchanges, which already involve an infinite number of operators, was obtained in \cite{Kulaxizi:2019tkd}. Based on such a computational scheme in holography, a recent paper \cite{Ceplak:2024bja} showed that the information about the black-hole singularity is encoded in the multi-stress tensor exchanges.\footnote{See also \cite{Louko:2000tp, Maldacena:2001kr, Kraus:2002iv, Fidkowski:2003nf, Festuccia:2005pi} for previous works on probing the black-hole singularity via conformal correlators.}   

We note that, in the geodesic limit, where the conformal dimension of the light scalars is large, using holography with a planar black hole, the work \cite{Parnachev:2020fna} obtained an exact scalar correlator  in $d=4$.  Utilising a connection with the Nekrasov-Shatashvili partition function \cite{Nekrasov:2009rc}, the work \cite{Dodelson:2022yvn} computed the exact $d=4$ thermal scalar correlator via holography with a Schwarzschild black hole.  Related holographic computations can be found in, $e.g.$, \cite{Alday:2020eua, Grinberg:2020fdj, Rodriguez-Gomez:2021pfh, Rodriguez-Gomez:2021mkk, Krishna:2021fus,  Parisini:2022wkb, Bajc:2022wws,  Dodelson:2023vrw,  Abajian:2023jye, Parisini:2023nbd, Fardelli:2024heb}. 

Here we are interested in the same CFT observable, but we would like to make progress towards computing the exact correlator that encapsulates all minimal-twist multi-stress tensor contributions at finite $\Delta$, without relying on a gravity computation. In addition, we would like to know how to resum all the contributions explicitly in the field theories.

Several CFT approaches have been developed to compute the $d>2$ multi-stress tensor contributions in an OPE expansion. Using the Lorentzian inversion formula \cite{Caron-Huot:2017vep, Simmons-Duffin:2017nub}, the $d=4$ double-stress tensor contributions were computed in \cite{Li:2019zba, Li:2020dqm}. The lightcone bootstrap was employed in \cite{Karlsson:2019dbd, Karlsson:2020ghx} to obtain the closed-form expressions for both double- and triple-stress tensor exchanges in $d=4$ by adopting a correlator  ansatz. More recently, refs. \cite{Parisini:2022wkb, Parisini:2023nbd} applied the ambient space approach \cite{FG1985, 2007arXiv0710.0919F} to compute the $d=4$ multi-stress tensor contributions. To our knowledge, the available results obtained via CFT and holography methods are mutually consistent, providing non-trivial cross-checks for each method. 

This note builds on the previous work  \cite{Huang:2023ikg}, which put forward a set of linear partial differential equations as the higher-dimensional generalisation of the $d=2$ null-state differential equations at large central charge.  In a near-lightcone expansion, the solutions to these equations are consistent with the available results pertaining to the contributions from the minimal-twist multi-stress tensors to the heavy-light correlator at large central charge. In this note, we discuss the exact solutions to these linear differential equations, assuming their validity.  The exact solutions then represent the resummation of all minimal-twist multi-stress tensors. %We hope that the new results obtained in this note will help clarify the validity of the proposed equations.  
For concreteness, we will focus on $d=4$ as the higher-dimensional example and compare our results with those obtained in the more familiar $d=2$ case.

A restriction of this type of linear differential equations is that the corresponding correlators at large central charge require the light scalars to have certain {\it negative} integer conformal dimensions. Specifically,  at large central charge $c$, the level-2 degenerate scalar in $d=2$ has the following conformal dimension
\begin{align}
 d=2: ~~~~~~~ h= {\Delta \over 2}= - {1\over 2} + {\cal O}\Big({1\over c}\Big) \ .
\end{align}  
This required value can be derived using the Virasoro algebra \cite{BELAVIN1984333}. While having a non-unitary value may seem like a significant limitation, the information about the Virasoro blocks at general $\Delta$ may be obtained by analytically continuing the correlators involving degenerate operators; see, $e.g.$, \cite{Chen:2016cms} for a related discussion. 

It remains an intriguing question whether analogous degenerate states exist in higher dimensions as a consequence of a symmetry structure hidden within a class (or a subsector) of higher-dimensional CFTs with a simple dual gravity description. Here, we shall simply assume that similar degenerate-like scalars exist, with conformal dimensions taking three special values at large central charge $C_T$
 \begin{align}
d=4: ~~~~~~~ \Delta &= -1, -2, -3  +  {\cal O}\Big({1\over C_T}\Big)  \ .
\end{align}   
It was observed that certain patterns of the $d=4$ thermal scalar correlator at large $C_T$ emerge at these three special values \cite{Fitzpatrick:2019zqz}. See also \cite{Fitzpatrick:2019efk} for a closely related discussion.  Furthermore, the work \cite{Fitzpatrick:2020yjb} noted that certain non-minimally coupled bulk interactions affect the $d=4$ near-lightcone correlator, but the corrections vanish at these three special values. While this note assumes CFTs with a simple gravity dual, it would be interesting to better understand why these special scalars do not ``feel" the non-minimally coupled bulk interactions and how generalisable this phenomenon is.  Moreover, it would be useful to identify physical quantities that are insensitive to the $\Delta$ dependence, which could be probed by these degenerate-like fields in higher dimensions.

At large central charge, the enhancement due to heavy scalars plays an important role. We take $C_T \to \infty$ and $\Delta_H \to \infty$, with the ratio $\Delta_H/C_T$ held fixed. 
We denote the (normalised) heavy-light scalar correlator in $d=2$ and $d=4$ as
\begin{align}
&d=2:~~~~ F(z)={\langle {\cal O}_H (\infty) {\cal O}_L(z) {\cal O}_L(0)  {\cal O}_H(1)\rangle \over   \langle {\cal O}_H(\infty)  {\cal O}_H (1) \rangle \langle {\cal O}_L(z) {\cal O}_L (0) \rangle}\Big|_{T^n} 
= \sum^{\infty}_{n=0} G_n (z) \eta^n   \\
\label{4dLCC}
&d=4:~~~~ F(z, \bar z)= \lim_{\bar z \to 0} {\langle {\cal O}_H (\infty) {\cal O}_L(1) {\cal O}_L(1-z, 1-\z)  {\cal O}_H(0)\rangle \over   \langle {\cal O}_H(\infty)  {\cal O}_H (0) \rangle \langle {\cal O}_L(1) {\cal O}_L (1-z, 1-\z) \rangle}\Big|_{T^n} 
= \sum^{\infty}_{n=0} G_n (z) (\mu \bar z)^n  
\end{align}   
where $T^n$ indicates the exchanged operators made out of $n$ stress tensors (and derivatives). 
The holomorphic $d=2$ correlator has an expansion in terms of $\eta \sim {\Delta_H\over  c}$ while the $d=4$ correlator has an expansion in terms of $\mu \sim {\Delta_H\over C_T}$.\footnote{This ratio is the coefficient of the first correction to the AdS bulk metric in a near-boundary expansion. In the convention of, $e.g.$, \cite{Fitzpatrick:2019zqz}, $\eta = {\Delta_H\over 2 c}$, $\mu = {160 \over 3}{\Delta_H\over C_T}$, where the proportionality constants will not be important in the analysis of this note.  In $d=2$, the (heavy-light) Virasoro vacuum block corresponds to summing over all orders of contributions in an expansion in terms of this ratio \cite{Fitzpatrick:2014vua, Fitzpatrick:2015zha}; see also, $e.g.$, \cite{Fitzpatrick:2015qma, Kulaxizi:2018dxo}.}  Note that the $d=4$ expression  assumes a near-lightcone limit which, in our convention, is $\z\to 0$. If one begins with the full $d=4$ correlator, one may perform an expansion in $\mu$ and subsequently an expansion in $\z$, retaining the leading $\z$ term. In both $d=2$ and $d=4$, the $\Delta$-dependent functions $G_n (z)$ satisfy $G_0 (z)= 1, G_{n>1}(0)=0$ -- these will be treated as boundary conditions of the differential equations we consider in what follows.

The parameters $\eta$ and $\mu$ are related to the Hawking temperature $T_H$ of the BTZ black hole and the AdS$_5$-Schwarzschild black hole, respectively, \cite{Witten:1998zw}: 
\begin{align}
&d=2:~~~~~ T_H= {1\over 2\pi }\sqrt{24 \eta  -1}  \\
&d=4:~~~~~ T_H= {1\over \sqrt{2} \pi} \sqrt{ { 4 \mu+1}\over {\sqrt{4 \mu+1} -1}}
\end{align} 
The AdS radius is set to 1. A planar black hole, $i.e.$, a black brane, is often adopted as a simplification which corresponds to the high-temperature limit in which $\mu$ approaches infinity. In this case, the relation between $\mu$ and $T_H$ is simplified to $\mu = (\pi T_H)^4$. In this work, we expect that our correlator results correspond to a holographic computation with a {\it spherical} black hole. In $d=4$, for simplicity, we will often absorb the parameter $\mu$ into $\z$ in the near-lightcone expression. 

We will consider a linear ordinary differential equation in $d=2$, which is the level-two null-state equation at large $c$ applied to the four-point scalar correlator, and three linear partial differential equations in $d=4$, proposed in \cite{Huang:2023ikg}. 
The equations that require $\Delta = -1$ will serve as our primary examples; they may be succinctly represented as
 \begin{align} 
&d=2: ~~~~  x^2  u_{xx} + c_2 u =0  \\
&d=4: ~~~~  x^3 u_{xxxy} + u=0 
\end{align}    
with $c_2= 6\eta $, $x=1-z$, $y= -\z$. We have rescaled $\z$ to absorb $\mu$ in $d=4$.   
The relations between the correlators and the functions $u(z)$ and $u(z, \z)$ in $d=2$ and $d=4$ are given by $u(z) = z F(z)$ and $u(z, \z) = (z^2 \z) F(z, \z)$, respectively. 
The exact solution to the $d=2$ equation corresponds to the large-$c$ Virasoro vacuum block, evaluated at the degenerate value $\Delta = -1$.  
The $d=4$ equation is straightforward to solve in a small $\z$ expansion using the boundary conditions mentioned above; these solutions are consistent with the available results computed via holography and CFT methods. In this note, we will discuss the exact solution. Similar $d=4$ equations for the $\Delta = -2$ and $\Delta = -3$ cases were also proposed in \cite{Huang:2023ikg}, and we will discuss their exact solutions as well.

The underlying reason why the $d=4$ large-$C_T$ correlator satisfies a set of linear differential equations at special $\Delta$ remains to be better understood. In this work, we assume the validity of these differential equations and focus on solving them. One of our primary motivations is to explore the implications of these differential equations as the new results may help test their validity and potentially shed light on the reason for their existence in higher dimensions. 

\subsection{Summary of the main results}  

Here we provide a brief summary of the simplest expressions from the $\Delta=-1$ case. Other cases will be discussed later in the text.

We adopt a convenient variable $v = \ln(1-z)$, which helps identify patterns in the solutions and simplifies various intermediate computations.
Using the differential equations, we find that, for any number $n$ of the multi-stress tensor exchanges, the scalar correlators can be written in terms of the modified Bessel function of the first kind:
\begin{align}
&d=2: ~~~~~ G^{{\Delta=-1}}_n(z)= - \frac{ (-6)^{n}  \sqrt{\pi }  }{\Gamma (n+1)}  {(1-z)^{1\over 2}\over z} ~  v^{n+\frac{1}{2}} I_{n+\frac{1}{2}}\left(\frac{v}{2}\right)\Big|_{v=\ln(1-z)} \\
&d=4: ~~~~~ G^{{\Delta=-1}}_n(z)=  \frac{2 \sqrt{\pi }}{z^2 \Gamma^2 (n+1) \Gamma (n+2) } \int_0^v d t (v-t)^n e^{\frac{3 t}{2}}  t^{n+\frac{1}{2}} I_{n+\frac{1}{2}} \big(\frac{t}{2}\big)\Big|_{v=\ln(1-z)}
\end{align} 
We will refer to these $n$-dependent expressions as ${\cal O}(n)$ correlators. Even in the $d=2$ case, the explicit $n$-dependent structure involving degenerate scalars, to our knowledge, has not been pointed out in the literature. Summing over $n$ in $d=2$ yields the large-$c$ Virasoro vacuum block,  evaluated at $\Delta=-1$.

A direct summation in $d=4$ is more challenging, but we find that the expression simplifies in the complex frequency domain, $i.e.$, the $s$-plane. 
We can express the exact, resummed correlator as an inverse Laplace transform. This method is also applicable to the $d=2$ case. The results are 
\begin{align}
&d=2: ~~~~~  F^{{\Delta=-1}}(z)= -{1\over z} \mathcal{L}^{-1}\Big\{ \frac{1}{s(s-1)+6 \eta}\Big\} (v)\Big|_{v=\ln(1-z)}  \\
&d=4: ~~~~~  F^{{\Delta=-1}}(z, \z)= {2 (1-z) \over z^2 \z}  ~ \mathcal{L}^{-1} \Big\{ e^{\frac{\z}{s (s^2 -1)}} \Big\} (v)\Big|_{v=\ln(1-z)}  
\end{align} 
In this $d=4$ expression, the parameter $\mu$ is absorbed into $\z$ and we assume $v\neq 0$.\footnote{The shifting property of the inverse Laplace transform can be used to adjust expressions in the $s$-plane, but this does not alter the results in position space.}  An interesting feature of the $d=4$ correlator is its presence of {\it essential singularities} in the $s$-plane. We also obtain the resumed expressions for the cases with $\Delta=-2$ and $\Delta=-3$ in $d=4$, which exhibit similar essential singularities.  

We will derive the above exact expressions using two distinct methods:  
(i) By examining correlator patterns in an expansion, we obtain the explicit ${\cal O}(n)$ expressions, then we sum over $n$; 
(ii) We solve the linear differential equations in the $s$-plane with suitable boundary conditions.  
The first method allows us to obtain the ${\cal O}(n)$ expressions, while the second method yields the exact results directly.  

Another independent observation based on the differential equations is that the correlators can be expressed as certain mode summations. In $d=2$, these modes correspond to the Virasoro modes $L_{m}$, which form a closed algebra. We find similar mode representations for the $d=4$ correlator. Denoting $\{k_i\}$ as a set of independent modes, we find that the contributions of multi-stress tensors at large central charge can be organised as follows: 
{\allowdisplaybreaks
\small
\begin{align}
&d=2:~~~~~ F^{{\Delta= -1}}(z)= 1+ \sum_{n=1}^{\infty} \sum_{\{k_i \}=2}^{\infty}  \Bigg[ \frac{ (-6)^n \prod _{i=1}^n (k_i-1)}{\prod _{l=1}^n \left(\sum _{i=1}^l k_i\right) \left(\sum _{i=1}^l k_i+1\right)} \eta^n z^{\sum _{i=1}^n k_i} \Bigg]\nn\\
&~~~~~~~~~~~~~~~~~~~~~~~~~~~~ =1  - \sum_{k_1=2}^{\infty}\frac{ 6 (k_1-1) z^{k_1}}{k_1 (k_1+1)} \eta \nn\\
  &~~~~~~~~~~~~~~~~~~~~~~~~~~~~~~~~~~~~~~ + \sum_{k_1, k_2=2}^{\infty}\frac{36 (k_1-1) (k_2-1) z^{k_1+k_2}}{k_1 (k_1+1) (k_1+k_2) (k_1+k_2+1)} \eta^2 + {\cal O} (\eta^3)  \\
\nn\\
&d=4:\nn\\
&  F^{{\Delta= -1}}(z,\z)=  1+ \sum_{n=1}^{\infty} \sum_{\{k_i \}=3}^{\infty}  \Bigg[ \frac{(-1)^n}{2^n \Gamma (n+2) } \frac{\prod _{i=1}^n \Big( (k_i-1) (k_i-2)\Big) }{\prod_{l=1}^n  \Big[ \left(\sum_{i=1}^l k_i\right) \left(\sum _{i=1}^l k_i+1\right) \left(\sum_{i=1}^l k_i+2\right)  \Big] } \z^n z^{\sum _{i=1}^n k_i}   \Bigg]\nn\\
& ~~~~~~~~~~~~~~~~ =1 - \sum_{k_1=3}^{\infty} \frac{(k_1-1) (k_1-2) z^{k_1} }{4 k_1 (k_1+1) (k_1+2)} \z \nn\\
&~~~~~~~~~~~~~~~~~~~~~~~~+ \sum_{k_1, k_2=3}^{\infty}\frac{ (k_1-1) (k_1-2) (k_2-1) (k_2-2)  z^{k_1+k_2} }{24 k_1 (k_1+1) (k_1+2) (k_1+k_2) (k_1+k_2+1) (k_1+k_2+2)}  \z^2 + {\cal O} (\z^3) 
\end{align}}Similar mode expressions for $\Delta= -2, -3$ are also worked out based on the corresponding $d=4$ differential equations.    
In $d=2$, the mode pattern can be organised in a diagrammatic way for any $\Delta$, as developed in \cite{Fitzpatrick:2015foa}.\footnote{An extension to the $d=2$ $\cal W$-vacuum blocks, determined by the higher-spin symmetries \cite{Z1985}, can be found in \cite{Karlsson:2021mgg}.} While we have not found the organising rules in $d=4$, we hope that these mode representations of the near-lightcone correlator will be helpful in developing an effective field theory in higher dimensions. 

The remainder of this note is organised as follows:
In Section \ref{sec2}, we use the level-2 null-state equation at large central charge to show that the  $d=2$ correlator in an expansion can be expressed as a Bessel function. 
We show the resummation gives the Virasoro vacuum block. We then point out a relation between the $d=2$ and $d=4$ correlators, both evaluated at $\Delta= -1$. This enables us to consider the  resummation in $d=4$.  In Section \ref{sec3}, we observe a simple pattern in the $s$-plane which facilitates the computation.   We also discuss the $\Delta=-2, -3$ cases in $d=4$.  In Section \ref{sec4}, we solve both the $d=2$ and $d=4$ equations directly in the $s$-plane and verify the resummation results.  Using the differential equations, in Section \ref{sec5} we show that the correlators can be expressed as mode summations.  In Section \ref{sec6}, we conclude with some future questions. Some explicit ${\cal O}(n)$ correlators are collected in Appendix \ref{AppA}, and we list the mode expressions in the first few orders in Appendix \ref{AppB}. 

\section{Multi-stress tensor exchanges as a Bessel function}
\label{sec2}

In this section, we begin by revisiting the more familiar context of $d=2$ CFTs at large central charge,   focusing on the heavy-light 4-point scalar correlator with the level-2 degenerate light scalars.  
We find that the contributions to the correlator from any number of the exchanged stress tensors can be written as a Bessel function. 
This expression is useful as it allows us to see the explicit $n$-dependence of the correlator, where $n$ is the power of the expansion parameter $\eta$. 
Furthermore, by focusing on the light scalars with $\Delta= -1$, we observe that the $d=4$ correlator in the near-lightcone expansion can be related to the $d=2$ correlator in a simple way.  Consequently, the contributions of the $d=4$ minimal-twist multi-stress tensors at ${\cal O}(n)$ are given by a Bessel function as well.  We present the $d=4$ resummation result as double integrals.  We will revisit the resummation structure in the next section, where we will also discuss the correlator with the other values of $\Delta$ in $d=4$. 

\subsection{$d=2$}

The level-2 null-state equation \cite{BELAVIN1984333}, when applied to the (holomorphic) heavy-light four-point correlator at large central charge, can be written as an ordinary differential equation: 
\begin{align}
\label{bpz}
\Big(\partial_z^2 +{ 6 \eta \over (1-z)^2} \Big) Q^{\Delta=-1}(z)=0   
\end{align}  
where $\eta = {h_H\over c}$= $\Delta_H\over 2c$.  The equation for $F^{\Delta=-1}(z) =  z^{-1} Q^{\Delta=-1}(z)$ is
\begin{align}
\label{2dF}
\Big(\partial^2_z + {2\over z}  \partial_z + {6 \eta\over (1-z)^2} \Big)  F^{\Delta=-1}(z)=0   \ . 
\end{align} 
Note the equation transforms covariantly under $z \to {z\over z-1}$, which corresponds to the crossing symmetry due to the exchange of two identical light scalars. 

We can solve the $d=2$ equation exactly, but instead, we will adopt an expansion in powers of $\eta$, which enables us to derive a closed-form ${\cal O}(n)$ expression valid at all orders. 
This approach will allow us to uncover the detailed structure of the correlator, and we will apply a similar strategy in $d=4$ later.

We write\footnote{In $d=4$, we write $F(z,\z) = \sum_{n=0}^{\infty} G_{n}(z) \z^n$.} 
\begin{align}
F(z) = \sum_{n=0}^{\infty} G_{n}(z) \eta^n = G_0(z) + G_1 (z)\eta+ G_2 (z)\eta^2+G_3 (z)\eta^3 +\cdots 
\end{align} 
and require 
\begin{align}
\label{BC}
 G_0(z)=1 \ , ~~~~ \lim_{z \to 0}G_i (z) =0  ~~ ({\rm for~} i > 0)  \ .
\end{align} 
The solutions satisfy a recursive relation \cite{Huang:2023ikg}
\begin{align}
\label{int2d}
G^{{\Delta=-1}}_n (z) = {1\over z} R^{{\Delta=-1}}_n(z)    \ , ~~ R^{{\Delta=-1}}_{n+1}(z)= - 6 \int^z_0 ds_2 \int^{s_2}_0 ds_1  ~\frac{R^{{\Delta=-1}}_n(s_1)}{(1-s_1)^2}  
\end{align} 
with $R^{{\Delta=-1}}_0(z)= z$ as the initial condition. The ${\cal O}(n)$ solutions correspond to $n$ numbers of stress-tensor exchanges.   

To identify a general patten, we adopt the following variable: 
\begin{align}
\label{vv}
v= \ln (1-z) \ . 
\end{align}  
The solutions to the $d=2$ differential equation in the first few orders can be written as 
{\allowdisplaybreaks
\small
\begin{align}
& \label{R02d} R^{{\Delta=-1}}_0(v)= 1-e^v \\
& \label{R12d}  R^{{\Delta=-1}}_1(v)= 6 \Big(v+2+e^v (v-2)\Big) \\
& \label{R22d}  R^{{\Delta=-1}}_2(v)=18 \Big(v^2+6 v+12-e^v \left(v^2-6 v+12\right)\Big)\\
& \label{R32d} R^{{\Delta=-1}}_3(v)=36 \Big(v^3+12 v^2+60 v+120+e^v \left(v^3-12 v^2+60 v-120\right)\Big)\\
& \label{R42d}  R^{{\Delta=-1}}_4(v) = 54 \Big(v^4+20 v^3+180 v^2+840 v+1680 -e^v \left(v^4-20 v^3+180 v^2-840 v+1680\right)\Big) 
\end{align}}Higher-order solutions can be worked out readily, and we collect more explicit expressions in Appendix \ref{AppA}.  

Let us make some observations on these $d=2$ solutions.

(i) The overall coefficients satisfy 
\begin{align}
\label{2doverall}
\frac{6^n}{\Gamma (n+1)} \ .
\end{align} 

(ii) The $e^v$ part of the solutions is completely fixed by the part not multiplied by $e^v$; one can write $R_n(v)$ as $A_n(v) - e^vA_n(-v)$. 

(iii) Focusing on the part not multiplied by $e^v$, the coefficients are listed below:
{\allowdisplaybreaks
\begin{align}
n=0: ~~~~~ &\{1\}\\
n=1: ~~~~~ &\{1, 2\}\\
n=2: ~~~~~ &\{1, 6, 12\}\\
n=3: ~~~~~ &\{1, 12, 60, 120\}\\
n=4: ~~~~~ &\{1, 20, 180, 840, 1680\}\\
n=5: ~~~~~ &\{1, 30, 420, 3360, 15120, 30240\}\\
n=6: ~~~~~ &\{1, 42, 840, 10080, 75600, 332640, 665280\}\\
n=7: ~~~~~ &\{1, 56, 1512, 25200, 277200, 1995840, 8648640, 17297280\} \ .
\end{align}}For a fixed $n$, the sequence can be organised by 
\begin{align}
\frac{(n+m)!}{(n-m)! m! } \ , ~~~~~~ m= 0,1,2,3,\dots, n \ .
\end{align} 
The  power of $v$ is given by $n-m$.  One may verify this pattern using higher-order terms.\footnote{This pattern appears as the sequence {\it A160481} in the Online Encyclopedia of Integer Sequence \cite{OEIS}.}  The ${\cal O}(n)$ correlator, $G_n(v)$, can be obtained by summing over $m$ with a range set by $n$.   
 
Putting everything together, including the part multiplied by $e^v$ and the overall coefficient \eqref{2doverall}, we obtain, to our knowledge, a new expression in $d=2$:
{\allowdisplaybreaks
\begin{align}
\label{2dOn}
G^{{\Delta=-1}}_n(v)= {1\over 1- e^{v}}R^{{\Delta=-1}}_n(v) &=  {1\over 1- e^{v}} \sum_{m=0}^n 
\frac{6^n }{\Gamma (n+1) } \frac{ (n+m)! }{(n-m)! m! } \Big( v^{n-m}  - e^v (-v)^{n-m}   \Big)\nn\\
&= \frac{ (-6)^{n}  \sqrt{\pi }  }{\Gamma (n+1)}  {e^{v\over 2}\over e^{v}-1} ~  v^{n+\frac{1}{2}} I_{n+\frac{1}{2}}\left(\frac{v}{2}\right)
\end{align}}where $I_{n+\frac{1}{2}}\left(\frac{v}{2}\right)$ is the modified Bessel function of the first kind.  

Therefore, the resumed $d=2$ correlator is 
{\allowdisplaybreaks
\begin{align}
F^{{\Delta=-1}}(v) 
&=  {\sqrt{\pi } v^{1\over 2}  e^{v\over 2}\over e^{v}-1}  \sum_{n=0}^{\infty}  \frac{ (-6)^{n}  }{\Gamma (n+1)}~  v^{n} I_{n+\frac{1}{2}}\left(\frac{v}{2}\right) \eta^n \\
&=  {\sqrt{\pi } v^{1\over 2}  e^{v\over 2}\over  e^{v}-1}  \sum_{n=0}^{\infty}  \frac{ (-24\eta )^{n} }{n!}~  ({v\over 4})^{n} I_{n+\frac{1}{2}}\left(\frac{v}{2}\right) 
\end{align}}where, in the second line, we have arranged the expression to fit the multiplication theorem of the Bessel function:
\begin{align}
{I_a (\lambda x)} = \lambda^{a}   \sum_{n=0}^{\infty} {(\lambda^2-1)^n\over n!} ({x\over 2})^n I_{n+a} (x) \ .
\end{align} 
Let $a={1\over 2}$, $x={v\over 2}$ and 
\begin{align}
 \lambda^2-1 = - 24\eta
\end{align}  
which has two solutions, $\lambda= \pm \sqrt{1-24 \eta }$. The different sign choices give the same $F(z)$. 
We obtain
\begin{align}
\label{VB1}
F^{{\Delta=-1}}(z)=   {\sqrt{\pi } v^{1\over 2}  e^{v\over 2}\over  e^{v}-1}  {I_{1\over 2} ( {1\over 2}\sqrt{1-24 \eta } ~v)\over ( \sqrt{1-24 \eta })^{{1\over 2}}}\Big|_{v=\ln(1-z)}  
&=  -\frac{ \sqrt{1-z}}{z}  \frac{\sinh \left(\frac{1}{2} \sqrt{1-24 \eta } \ln(1-z)\right)}{{1\over 2}\sqrt{1-24 \eta }} \ . 
\end{align} 
This reproduces the $d=2$ Virasoro vacuum block at large $c$ obtained in \cite{Fitzpatrick:2014vua, Fitzpatrick:2015zha}. (For $\eta > {1\over 24}$, the sinh becomes a sine.)   As we start with the null-state differential equation, this correlator result requires the level-two degenerate scalars.

In this work, we focus on the $\Delta = -1$ case in $d=2$, and the $\Delta = -1, -2, -3$ cases in $d=4$. To study from a first principle the $d=2$ correlators involving higher-level degenerate scalars, one needs to consider other null-state equations. We will not discuss these higher-order differential equations in $d=2$. 

\subsection{$d=4$} 

In \cite{Huang:2023ikg}, three linear partial differential equations were proposed as the governing equations for the heavy-light scalar correlator in the near-lightcone regime in $d=4$ CFTs with a simple gravity dual description. These  equations are restricted to specific values of $\Delta$. 
Despite this, we hope that the analysis presented in this note could provide further insights into the analytic structure of a more general $d=4$ correlator at different values of $\Delta$.

The simplest case requires $\Delta= -1$. The equation is \cite{Huang:2023ikg}
\begin{align}
\label{4deq1}
\Big[\big(1+ \z \partial_{\z} \big) \big(  \partial^3_z + {6 \over z} \partial^2_z+{6\over z^2} \partial_z \big)+ {\mu \z \over (1-z)^3} \Big] F^{\Delta=-1}(z,\z) = 0 \ .
\end{align}   
In this expression, we adopt a conformal frame in which all four scalars live on a two-dimensional plane, and we keep the explicit $\mu$ dependence; in what follows, we will absorb $\mu$ into $\z$. 
The equation transforms covariantly under the crossing symmetry due to the presence of two identical light scalars. This is a Fuchsian-type equation, with factorised $z$- and $\z$-derivative parts. Denote 
\begin{align}
D^k_z = (z \partial_{z})^k \ , ~~~D^k_{\z} = ({\z} \partial_{\z})^k \ .
\end{align}
 Equation \eqref{4deq1} can be represented as
\begin{align}
\label{4deq1new}
\Big[\big(1+D_{\z} \big) \big(  D^3_z+ {6 }   D^2_z+{6 } D_z\big)+ {z^3 \z \over (1-z)^3} \Big]F^{\Delta=-1}(z,\z) = 0 \ .
\end{align} 
As mentioned in \cite{Huang:2023ikg}, a concise form of the equation is 
\begin{align}
\label{4dQeq1}
\Big(\partial_{\bar z} \partial^3_z  + {1 \over (1-z)^3} \Big)  Q^{\Delta=-1} (z,\bar z)=0 
\end{align}  
where $Q^{\Delta=-1}(z,\z) = z^2 \z F^{\Delta=-1}(z,\z)$. 
The equation can be solved order by order in a small $\z$ expansion using the boundary conditions \eqref{BC}. 
In such an expansion, the solutions satisfy a recursion relation \cite{Huang:2023ikg}
\begin{align}
\label{Recur4d1}
 Q^{\Delta=-1}(z,\z)&= \sum_{n=0} R^{\Delta=-1}_n(z) \z^{n+1}  \ , \nn\\
 R^{\Delta=-1}_{n+1}(z)&=  {-1\over n+2} \int^z_0 ds_3 \int^{s_3}_0 ds_2 \int^{s_2}_0 ds_1  ~\frac{R^{\Delta=-1}_n(s_1)}{(1-s_1)^3} 
\end{align}   with the initial condition $R^{\Delta=-1}_0(z)= z^2$.  
For $n>1$, the corresponding solution contains the contributions from an infinite number of primary operators built from stress tensors and derivatives. 

Given the $d=2$ ${\cal O}(n)$ expression in Section \ref{sec2}, let us try to find the corresponding  expression in $d=4$. 
Using the $v$-variable, the $d=4$ solutions in the near-lightcone expansion are given by 
{\small
\allowdisplaybreaks
\begin{align}
& \label{R0} R^{{\Delta=-1}}_0(v)= (1-e^v)^2 \\
&\label{R1}  R^{{\Delta=-1}}_1(v)= \frac{1}{4}  \left(v+3+e^{2 v} (v-3)+4 e^v v\right)\\
&\label{R2}  R^{{\Delta=-1}}_2(v)= \frac{1}{48} \left(v^2+9 v+24-8 e^v \left(v^2+6\right)+e^{2 v} \left(v^2-9 v+24\right)\right)\\
& \label{R3} R^{{\Delta=-1}}_3(v)= \frac{1}{1152} \Big(v^3+18 v^2+123 v+315\nn\\
&~~~~~~~~~~~~~~~~~~~~~~~~~~~~~~~~~~~~~~ +16 e^v \left(v^2+24\right) v +e^{2 v} \left(v^3-18 v^2+123 v-315\right)\Big)\\
& \label{R4} R^{{\Delta=-1}}_4(v) = \frac{1}{46080} \Big(v^4+30 v^3+375 v^2+2295 v+5760 \nn\\
&~~~~~~~~~~~~~~~~~~~~~~~~~~~ -32 e^v \left(v^4+60 v^2+360\right) +e^{2 v} \left(v^4-30 v^3+375 v^2-2295 v+5760\right)\Big)\ .
\end{align}}We collect more expressions in Appendix \ref{AppA}.  

The pattern is less transparent compared with the $d=2$ case, but we observe a connection between the $d=4$ and $d=2$ correlators in the similar expansion: 
\begin{align}
\label{4d2drelation}
\partial^{n+1}_v R^{{\Delta=-1}}_n(v)|_{d=4}= \frac{(-1)^{n+1}}{2^{n-1} 3^n\Gamma (n+2)}  e^v R^{{\Delta=-1}}_n(v)|_{d=2} \ .
\end{align}  
The functions $ R^{{\Delta=-1}}_n(v)|_{d=2}$ were discussed in Section \ref{sec2}. It is straightforward to verify this relation to high orders.  

Using the relation \eqref{4d2drelation} and the ${\cal O}(n)$ expression of the $d=2$ function $R^{{\Delta=-1}}_n(v)$ obtained in \eqref{2dOn}, we obtain the corresponding expression in $d=4$
\begin{align}
\label{4dOn1}
G^{{\Delta=-1}}_n(v)|_{d=4} &={1\over (1-e^{v})^2}R^{{\Delta=-1}}_n(v)|_{d=4} \nn\\
&=  \frac{2 \sqrt{\pi }}{(1-e^{v})^2 \Gamma^2 (n+1) \Gamma (n+2) } \int_0^v d t (v-t)^n e^{\frac{3 t}{2}}  t^{n+\frac{1}{2}} I_{n+\frac{1}{2}} \big(\frac{t}{2}\big) \ .
\end{align}  
To perform the $(n+1)$-fold integrals, we have used the Cauchy reduction formula
\begin{align} 
\label{cauchy}
\int_{0}^{x}{\cdots{\int_{0}^{x_{3}}{\int_{0}^{x_{2}}{A(x_1)\,dx_1\cdots dx_k}}}}
=\frac{1}{(k-1)!}{\int_{0}^{x}{(x-t)^{k-1}A(t)\,dt}}  \  .
\end{align} 
Note that $\partial^{(k)}_v R^{{\Delta=-1}}_n(v)\big|_{v=0, d=4}=0 $ for  $k=n, n-1, n-2, \dots, 0$.  

To sum over $n$, we may adopt the following integral representation of the Bessel function:
\begin{align} 
I_{a }(x)=\frac{({x\over 2})^{a } }{\sqrt{\pi } \Gamma \left(a +\frac{1}{2}\right)} \int_{-1}^1 \left(1-\sigma^2\right)^{a -\frac{1}{2}} e^{-\sigma x} \, d\sigma \ . 
\end{align} 
Setting $a= n+ {1\over 2}$, $x={t\over 2}$, using the expression \eqref{4dOn1}  we obtain a resummed $d=4$ correlator
\begin{align} 
\label{4dexact1}
F^{{\Delta=-1}}(v,\z)|_{d=4}& = \sum_{n=0}^{\infty}\z^n G^{{\Delta=-1}}_n(v)|_{d=4} \\
&=\frac{1}{(1-e^v)^2}\int_0^v d t 
\int_{-1}^{1} d\sigma~\Big[ t e^{-\frac{t}{2} (\sigma-3) }  \, _0F_3\big(;1,1,2;\frac{t^2(t-v) (\sigma^2-1)   \z}{4} \big)\Big] \ . \nn
\end{align}
In a small $\z$ expansion, this gives the first few orders of $R^{{\Delta=-1}}_n(v)|_{d=4}$ listed in Appendix \ref{AppA}. 

\section{Patterns in the $s$-plane}
\label{sec3}

The $d=4$ resummed correlator \eqref{4dexact1} is more complex than the $d=2$ Virasoro vacuum block at large central charge \eqref{VB1}.
However, we observe that the $d=4$ structure simplifies in the complex frequency domain.  

Let $s$ be a complex variable. Define the Laplace transform as 
\begin{align} 
A(s)= \mathcal{L} \{A(v)\}(s)= \int_0^{\infty} A(v) e^{-s v}  dv  \ .
\end{align} 
The inverse transform is 
\begin{align} 
A(v)= \mathcal{L}^{-1} \{A(s)\}(v)=  {1\over 2 \pi i}  \int_{\gamma- i \infty}^{\gamma +i\infty} A(s) e^{s v}  ds
\end{align}
where a constant $\gamma > 0$ is chosen so that the integration contour lies to the right of all singularities in $A(s)$. 
Transforming $R^{{\Delta=-1}}_n(v)|_{d=4}$ obtained in \eqref{4dOn1} to the $s$-plane gives
\begin{align} 
\label{R4d1}
R^{{\Delta=-1}}_n(s)|_{d=4} &= \frac{2}{\Gamma (n+2)  }  \Big(\frac{1}{ (s-2) (s-1) s}\Big)^{n+1} \ .
\end{align} 
Performing an inverse transform leads to 
\begin{align} 
 R^{{\Delta=-1}}_n(v)|_{d=4} 
&=\frac{2 e^v v^{3 n+2} }{\Gamma (n+2) \Gamma (3 n+3)} \, _1F_2\left(n+1;\frac{3 n}{2}+\frac{3}{2},\frac{3 n}{2}+2;\frac{v^2}{4}\right) \ .
\end{align} 
The function $G_n(v)$ can be obtained via $G_n(v)= {(1-e^v)^{-2}} R_n(v)$ in this $d=4$ case. 

We may sum over $n$ directly in the $s$-plane. Using \eqref{R4d1}, we obtain a simple form
\begin{align} 
 \sum_{n=0}^{\infty} \z^n R^{{\Delta=-1}}_n(s)|_{d=4} = {2\over  \z} \Big(e^{\frac{\z}{(s-2) (s-1) s}}-1\Big) \ .
\end{align}  
The correlator $F^{{\Delta=-1}}(v, \z)|_{d=4}$ then can be expressed as an inverse transform   
\begin{align} 
\label{4Dexact1}
F^{{\Delta=-1}}(v, \z)|_{d=4} = {2\over  (1-e^{v})^2  \z}~ \mathcal{L}^{-1} \Big\{e^{\frac{\z}{(s-2) (s-1) s}}-1\Big\}(v) \ . 
\end{align} 
This is one of our main results. It is remarkable that such a simple structure encodes the complete information about all minimal-twist multi-stress tensors probed by the light scalars with $\Delta= -1$. This structure effectively encapsulates the complexity of minimal-twist multi-graviton interactions in the dual gravity

We may further simplify the expression \eqref{4Dexact1} by noting that $\mathcal{L}^{-1} \{ -1 \} = - \delta (v)$ can be dropped for $v\neq 0$ ($i.e.$ $z \neq 0$). 
In addition, recall the shifting property of the inverse Laplace transform
\begin{align} 
\label{shift}
 \mathcal{L}^{-1} \{A(s+a) \} (v)= e^{ - a v } A(v) \ .
\end{align}  
By shifting $s \to s +1$ and focusing on the correlator at $v\neq 0$, we write 
\begin{align} 
\label{s12new}
 {2 e^{v}\over (1-e^{v})^2 \z}  ~ \mathcal{L}^{-1} \Big\{ e^{\frac{\z}{s (s^2 -1)}} \Big\} (v) \ . 
\end{align} 
We remark that the previous expression \eqref{4Dexact1} in a small $\z$ expansion does not produce a $\delta(v)\over \z$ term, 
as it is cancelled by the constant, $-1$, piece. Different expressions that can be related by \eqref{shift} should be considered equivalent. 

Because of the essential singularities from the exponential term, the inverse Laplace transform of the resummed $d=4$ correlator in the $s$-plane is challenging to carry out.
However, it is straightforward to perform the inverse transform in a small $\z$ expansion, giving the results that are consistent with the previously listed functions $R^{{\Delta=-1}}_n(v)|_{d=4}$.

Several immediate questions arise: How does the $d=2$ correlator look like in the $s$-plane? 
And, how about the $d=4$ correlator at other values of $\Delta$? 
We will discuss these questions in what follows. In particular, we will find that essential singularities are absent in the $d=2$ case, and we will see that these singularities arise due to the additional $\z$-derivative in the partial differential equations in higher dimensions.

\subsection{$d=2$}

Motivated by the above observation, let us revisit the $d=2$ structure.  
Focusing on the level-2 degenerate scalar, the $d=2$ solutions \eqref{R02d}-\eqref{R42d} (see also Appendix \ref{AppA}) can be organised in a simple way in the $s$-plane:
\begin{align} 
R^{{\Delta=-1}}_n(s)\big|_{d=2}=  6^n  \big( \frac{  -1 }{ s (s-1) } \big)^{n+1} \ .
\end{align} 
The inverse transform gives the Bessel function \eqref{2dOn}. Summing over $n$ in the $s$-plane gives 
\begin{align} 
\label{2dLap}
\sum _{n=0}^{\infty} \eta ^n R^{{\Delta=-1}}_n(s)\big|_{d=2}=  \frac{-1}{s(s-1)+6 \eta } \ .
\end{align} 
The inverse transform gives the Virasoro vacuum block at large central charge \eqref{VB1}, when  multiplied by the overall factor ${(1-e^v)}= z$.  

\subsection{$d=4$}

The $\Delta= -1$ case was discussed and now we consider the $\Delta = - 2, - 3$ cases. 
We assume that the scalars with $\Delta$ at these three values are at the ``same level" in $d=4$ at large $C_T$. 

\subsubsection*{$\Delta=-2$ case}

The corresponding Fuchsian-type linear differential equation is \cite{Huang:2023ikg}
{\allowdisplaybreaks
\begin{align}
&\Big[\Big( 2+\z \partial_{\z}\Big) \Big( \partial_z^4 + {12\over z}  \partial_z^3 +{36\over z^2}  \partial_z^2 +{24 \over z^3}  \partial_z \Big) \nn\\
&~~~~~~~~~~~~~~~~~~~~~~~~~ +\frac{6 \mu \z }{(1-z)^3}\Big( \partial_z+ \frac{3 (2-z)}{2  z(1-z) } \Big) \Big] F^{{\Delta=-2}}(z,\z) =0 
\end{align}}which can be written as 
{\allowdisplaybreaks
\begin{align}
&\Big[\Big(2+D_{\z}\Big) \left( D^4_z + {12} D^3_z+{36} D^2_z + {24}  D_z \right)  \nn\\
&~~~~~~~~~~~~~~~~~~~~~~~~~~~~ +\frac{6  z^3 \z }{(1-z)^3}\Big( D_z+ \frac{3 (2-z)}{2  (1-z) } \Big) \Big] F^{{\Delta=-2}}(z,\z) =0 
\end{align}}where, again, we have rescaled $\z$ to absorb $\mu$. 
The equation transforms covariantly under the exchange of two identical light scalars.  
A simpler form is \cite{Huang:2023ikg}
\begin{align}
\label{4dQeq2}
&\Big(\partial_{\z} \partial^4_z + \frac{6}{(1-z)^3} \partial_z+  \frac{ 9}{(1-z)^4} \Big) Q^{\Delta=-2}(z,\z)= 0 
\end{align} 
where $Q^{\Delta=-2}(z,\z) = z^3 \z^2 F^{\Delta=-2}(z,\z)$.
The boundary conditions in a small $\z$ expansion are given in \eqref{BC}; the solutions satisfy a recursion relation \cite{Huang:2023ikg}
{\allowdisplaybreaks
\begin{align}
\label{Recur4d2}
& Q^{\Delta=-2}(z,\z)= \sum_{n=0} R^{\Delta=-2}_n(z) \z^{n+2} \ , \nn\\ 
& R^{\Delta=-2}_{n+1}(z)=  {-1\over n+3} \int^z_0 ds_4 \int^{s_4}_0 ds_3 \int^{s_3}_0 ds_2 \int^{s_2}_0 ds_1 \Big( \frac{ 9 R^{\Delta=-2}_n(s_1)}{(1-s_1)^4}+ \frac{ 6 \partial_{s_1}R^{\Delta=-2}_n(s_1)}{(1-s_1)^3}  \Big)
\end{align}}with the initial condition $R^{\Delta=-2}_0(z)= z^3$. 

The solutions in the first few orders are listed in Appendix \ref{AppA}. 
We find that these solutions can be organised in a simple way in the $s$-plane: 
\begin{align}
R^{{\Delta=-2}}_n(v)= - \frac{2^{n+2} 3^{n+1} e^{\frac{3 v}{2}}  }{\Gamma (n+3)} \mathcal{L}^{-1} \Big\{ \frac{s^n}{ \left( (s^2-\frac{9}{4})  (s^2-\frac{1}{4}) \right)^{n+1}}\Big\} (v) \ .
\end{align}  We have used the shifting property of the inverse transform to adjust the expression. 
Summing over $n$ gives 
\begin{align}
\label{4Dexact2}
F^{{\Delta=-2}}(v,\z)&= {1\over (1-e^v)^3   }\sum_{n=0}^{\infty}R^{{\Delta=-2}}_n(v) \z^n\nn\\
&= \frac{e^{\frac{3 v}{2}}}{48  \left(1-e^v\right)^3 \z^2} \mathcal{L}^{-1} \Big\{9 s^{-2} +8 s^{-1} \left(2 s^3-5 s+12 \z\right)\nn\\
&~~~~~~~~~~~~~~~~~~~~~~~~~~~~~~~~~ -16 s^{-2} (s^2-\frac{9}{4})  (s^2-\frac{1}{4}) e^{\frac{6 s \z}{ (s^2-\frac{9}{4})  (s^2-\frac{1}{4}) }}\Big\} (v) \ .
\end{align} 
In a small $\z$ expansion this result reproduces $R^{{\Delta=-2}}_{n}(v)$ listed in Appendix \ref{AppA}. 
Like the $\Delta=-1$ case in $d=4$, the resummation at $\Delta= - 2$ also produces essential singularities.

We may simplify the expression by dropping the contact terms $\sim \delta(v)$ in the part that does not involve the essential singularities and write  
\begin{align}
F^{\Delta=-2}(v,\z)&=  \frac{e^{\frac{3 v}{2}}}{48  \left(1-e^v\right)^3  \z^2}    \Bigg(   9 v+96 \z  - 16 \mathcal{L}^{-1} \Big\{  s^{-2}(s^2-\frac{9}{4})  (s^2-\frac{1}{4}) e^{\frac{6 s \z}{ (s^2-\frac{9}{4})  (s^2-\frac{1}{4}) }} \Big\} (v)    \Bigg)  \ .
\end{align} 
Note that the role of the part that does not involve the essential singularities is to cancel the divergences at orders $1/\z^{2}$ and ${1/\z}$ in a small $\z$ expansion. 
Thus, the relevant correlator structure is fully encoded in the term involving the essential singularities. 

\subsubsection*{$\Delta=-3$ case}

The corresponding $d=4$ equation is \cite{Huang:2023ikg}
{\allowdisplaybreaks
\begin{align}
&\Big[\left(3+\z \partial_{\z} \right) \left( \partial^5_z + \frac{20}{z} \partial^4_z +\frac{120}{z^2} \partial^3_z  +\frac{240}{z^3} \partial^2_z +
\frac{120}{z^4} \partial_z  \right)  \nn\\
&~~~~~ + {21 \mu \z \over (1-z)^3} \Big( \partial_z^2 +\frac{8-5 z }{(1-z) z}  \partial_z + \frac{24 z^2-84 z+84 }{7 (1-z)^2 z^2} \Big) \Big] F^{\Delta=-3}(z,\z) = 0 \ .
\end{align}}We can express the equation as 
{\allowdisplaybreaks
\begin{align}
&\Big[\left(3+D_{\z} \right) \left( D^5_z + 20 D^4_z + 120 D^3_z  + 240 D^2_z + 120 D_z  \right)  \nn\\
&~~~~ + {21 z^3 \z \over (1-z)^3} \Big( D^2_z +\frac{8-5 z }{(1-z)} D_z + \frac{24 z^2-84 z+84 }{7 (1-z)^2 } \Big) \Big] F^{\Delta=-3}(z,\z) = 0 
\end{align}}where the parameter $\mu$ has been absorbed into $\z$.\footnote{In passing, we note that, using $D^k_z= (z \partial_z)^k$ and $D^k_{\z}=(\z \partial_{\z})^k$,  the resulting $d=4$ differential equations for all $\Delta=-1,-2,-3$ cases have the remaining $\z$-dependent coefficients only via $z^3 \z$.}
A compact expression is \cite{Huang:2023ikg}
\begin{align}
\label{4dQeq3}
\Big(\partial_{\z} \partial^5_z + \frac{21}{(1-z)^3} \partial^2_z+  \frac{63}{(1-z)^4} \partial_z+ \frac{72}{(1-z)^5} \Big) Q^{\Delta=-3}(z,\z) = 0 
\end{align} 
where $Q^{\Delta=-3}(z,\z)= z^4 \z^3 F^{\Delta=-3}(z,\z)$. In a small $\z$ expansion, the solutions between different orders are related to each other through a recursion relation \cite{Huang:2023ikg}
{\allowdisplaybreaks
\begin{align}
\label{Recur4d3}
&Q^{{\Delta=-3}}(z,\z)= \sum_{n=0} R^{{\Delta=-3}}_n(z) \z^{n+3}   \ , \nn \\
&R^{{\Delta=-3}}_{n+1}(z)=  {-1\over n+4}  \int^{z}_0 ds_5\int^{s_5}_0 ds_4 \int^{s_4}_0 ds_3 \int^{s_3}_0 ds_2 \int^{s_2}_0 ds_1  ~ \nn\\
& ~~~~~~~~~~~~~~\times \Bigg[ \frac{ 72 R^{{\Delta=-3}}_n(s_1)}{(1-s_1)^5} + \frac{ 63 \partial_{s_1}R^{{\Delta=-3}}_n(s_1)}{(1-s_1)^4}  + \frac{ 21 \partial^2_{s_1}R^{{\Delta=-3}}_n(s_1)}{(1-s_1)^3}\Bigg]  
\end{align}}with $R^{{\Delta=-3}}_0(z)= z^4$.  

The solutions in the first few orders are collected in Appendix \ref{AppA}.  
We find that these solutions can be organised in a simple way:
\begin{align}
R^{{\Delta=-3}}_n(v)= {16  e^{2 v}  3^{n+2} \over \Gamma (n+4) } 
\mathcal{L}^{-1}  \Big\{ \frac{(7 s^2-4)^n}{ \big( s (s^2-1)(s^2-4)\big)^{n+1} } \Big\}(v) \ .
\end{align} 
The shifting property of the inverse transform has been used to simplify the expression.  

Summing over $n$ leads to 
\begin{align}
\label{4Dexact3}
F^{{\Delta=-3}}(v,\z)&= {1\over (1-e^v)^4 } \sum_{n=0}^{\infty}R^{{\Delta=-3}}_n(v) \z^n\nn\\
&= \frac{e^{2 v}}{ \left(1-e^v\right)^4 \z^3}  \mathcal{L}^{-1}  \Big\{ \frac{8}{3 \left(7 s^2-4\right)^3}  \Big( 2 s^2 (s^2-1)^2 (s^2-4)^2 e^{\frac{3 (7 s^2-4) \z}{s(s^2-1)(s^2-4)}}\nn\\
&~~~~~~~~~~~~~~~~~~~~~~~~~~~~~~~~~~~~~~~~~~~~~ -9 \left(7 s^2-4 \right)^2 \z^2\nn\\
&~~~~~~~~~~~~~~~~~~~~~~~~~~~~~~~~~~~~~~~~~~~~~  -2 s^2 (s^2-1)^2 (s^2-4)^2\nn\\
&~~~~~~~~~~~~~~~~~~~~~~~~~~~~~~~~~~~~~~~~~~~~~ - 6 s (7 s^2-4) (s^2-1) (s^2-4) \z \Big) \Big\}(v) \ .
\end{align} 
Essential singularities again appear.  
Dropping the contact terms gives  
\begin{align}
\label{Fexact3}
&F^{{\Delta=-3}}(v,\z)= 
\frac{e^{2 v}}{ \left(1-e^v\right)^4 \z^3}  
\Bigg[ \frac{1}{823543}\Bigg( 3024  (22 v+343 \z) \cosh (\frac{2 v}{\sqrt{7}})\\
& -12 \sqrt{7} \Big((12 v+343 \z)^2+2156\Big) \sinh (\frac{2 v}{\sqrt{7}})\Bigg)  +\frac{16}{3}  \mathcal{L}^{-1}  \Big\{ \frac{s^2 (s^2-1)^2(s^2-4)^2 }{ \left(7 s^2-4\right)^3}   e^{\frac{3(7 s^2-4) \z}{ s (s^2-1)(s^2-4)}}   \Big\}(v) \Bigg] \ . \nn
\end{align} 
In a small $\z$ expansion this result reproduces $R^{{\Delta=-3}}_{n}(v)$ in Appendix \ref{AppA}.  

We remark that the role of the $\cosh$ and $ \sinh$ pieces in \eqref{Fexact3} is to cancel the divergences at orders ${1/\z^{3}}, {1/\z^{2}}$, and ${1/\z}$ in a small $\z$ expansion. 
Therefore, like the $\Delta=-1, -2$ cases, the information of minimal-twist multi-stress tensor exchanges is completely encoded in the term involving the essential singularities.  

\section{Solve the equations directly in the $s$-plane}
\label{sec4}

In this section, we solve the linear differential equations directly in the $s$-plane, in both $d=2$ and $d=4$, with suitable boundary conditions. 
While this approach does not allow us to obtain explicit ${\cal O}(n)$ expressions for the correlators, it provides a streamlined computation for verifying the exact, resummed correlators relevant to the 
 minimal-twist multi-stress tensor exchanges at large central charge.  

\subsection{$d=2$}

We start with the $d=2$ case, again focusing on the simplest, level-2 null-state equation at large $c$.  
Using the $v$-variable, the equation \eqref{bpz} can be written as 
\begin{align}
\Big(\partial^2_v-\partial_v+6 \eta \Big) Q(v)=0 \ .
\end{align}   We drop the superscript $\Delta= -1$.
 In this form, the coefficients are constants.  
Transforming the equation to the $s$-plane, we find
\begin{align}
\Big(s(s-1)+6 \eta \Big) \mathcal{L} \{ Q(v) \}(s)= (s-1) Q(0)+Q'(0)
\end{align}  
where $Q(0)$ and $Q'(0)= \partial_v Q(v)|_{v=0}$ are evaluated in position space. 
To be consistent with the boundary conditions \eqref{BC}, we impose $Q(0)=0$ and $Q'(0)= - 1$, which do not refer to a small $\eta$ limit. 
The equation then reduces to
\begin{align}
\Big(s (s-1)+6 \eta \Big) \mathcal{L} \{ Q(v) \}(s)= -1
\end{align}  
which gives  
\begin{align}
Q(v)= \mathcal{L}^{-1}\Big\{ \frac{-1}{s(s-1)+6 \eta}\Big\} (v)= \frac{- e^{v\over2} \sinh \left(\frac{1}{2} \sqrt{1-24 \eta } v\right)}{\frac{1}{2}  \sqrt{1-24 \eta }} \ .
\end{align}  
Recall $Q(v)= (1-e^{v}) F(v)$ in the $d=2$ case. 
The result is consistent with \eqref{2dLap}, reproducing the large-$c$ Virasoro vacuum block \eqref{VB1} involving the level-two degenerate scalars, which we previously obtained through an explicit summation.

\subsection{$d=4$}

\subsubsection*{$\Delta= -1$ case}

The $d=4$ equation \eqref{4dQeq1} can be written as 
\begin{align}
\Big(\partial^3_v \partial_{\z}-3  \partial^2_v \partial_{\z} +2 \partial_v \partial_{\z} -1 \Big)Q(v,\z) = 0 
\end{align} 
where the coefficients are independent of coordinates. Again, the $\Delta= -1$ superscript is dropped.
Transforming the equation to the $s$-plane gives
\begin{align}
&\mathcal{L} \{V(v,\z)\}(s)-s (s-1) (s-2) \partial_{\z} \mathcal{L} \{ Q(v,\z) \}(s) \nn\\
&~~~~~~~~~~~~~~ +(s-1) (s-2) Q^{(0,1)}(0,\z)+(s-3) Q^{(1,1)}(0,\z)+Q^{(2,1)}(0,\z) = 0 \ .
\end{align} 
We impose $Q^{(2,1)}(0,\z)= 2$ and $Q^{(1,1)}(0,\z)=Q^{(0,1)}(0,\z)=0$ to be consistent with the boundary conditions \eqref{BC}. 
Here the more general boundary conditions are imposed without requiring a small $\z$ expansion. 
The equation becomes 
\begin{align}
\mathcal{L} \{Q(v,\z)\}(s)-s (s-1) (s-2) \partial_{\z} \mathcal{L} \{ Q(v,\z) \}(s) +2 = 0 \ .
\end{align} 
Therefore, we have 
\begin{align}
\mathcal{L} \{Q(v,\z)\}(s)= c_1(s) e^{\frac{\z}{s (s-1) (s-2)}}  -2  \ .
\end{align}  
To determine the function $c_1(s)$, we consider a small $\z$ expansion: 
\begin{align}
 \lim_{\z \to 0} \mathcal{L} \{ (1-e^v)^2 F(v,\z)\}(s) = \lim_{\z \to 0} \mathcal{L} \{{Q(v,\z) \over  \z}\}(s) =\frac{c_1(s)-2}{\z} + {\cal O} (\z^0)  
\end{align}   
where we have included the relative factor $z^2 \z$ between the functions $F$ and $Q$. 
Setting $c_1(s)=2$ to remove the divergence, we obtain
\begin{align}
 F(v,\z)=   {1\over (1-e^v)^2 \z}  Q(v,\z) ={2\over (1-e^v)^2 \z}  \mathcal{L}^{-1}\Big\{ e^{\frac{\z}{s (s-1) (s-2)}}  -1 \Big\} (v)   \ .
\end{align} 
This agrees with the result \eqref{4Dexact1} obtained through a summation.   

As a byproduct of this computation, we see that the essential singularities from exp$(\frac{\z}{s(s-1)(s-2)})$ originate from the $\z$-derivative part of the $d=4$ partial differential equation. (This is also the case when $\Delta= -2, -3$, which we discuss next.)  The corresponding $d=2$ equation is an ordinary differential equation and essential singularities are absent. 

\subsubsection*{$\Delta= -2$ case}

We write the corresponding $d=4$ equation \eqref{4dQeq2} as
\begin{align}
\Big(\partial^4_v \partial_{\z}-6 \partial^3_v \partial_{\z} +11 \partial^2_v \partial_{\z}-6 \partial_v \partial_{\z} -6 \partial_v +9 \Big)Q(v,\z) = 0
\end{align} 
where the coefficients are constants and we drop the superscript. 
In the $s$-plane, we have  
{\allowdisplaybreaks
\begin{align}
&(9-6 s) \mathcal{L} \{ Q(v,\z) \}(s)  +s (s-1) (s-2) (s-3)\partial_{\z} \mathcal{L} \{ Q(v,\z) \}(s)\nn\\
&~~~~ -Q^{(3,1)}(0,\z) -(s-6) Q^{(2,1)}(0,\z)-(s-1)(s-2)(s-3) Q^{(0,1)}(0,\z)\nn\\
&~~~~~~~~~ - \big(s (s-6) +11 \big)Q^{(1,1)}(0,\z)+6 Q(0,\z) = 0 \ .
\end{align}}Requiring in this case $Q^{(3,1)}(0,\z) = -12 \z$, $Q(0,\z)=Q^{(0,1)}(0,\z)=Q^{(1,1)}(0,\z)=Q^{(2,1)}(0,\z)=0$, which are motivated by the boundary conditions \eqref{BC} on the function $F$, the equation reduces to 
\begin{align}
(9-6 s) \mathcal{L} \{ Q(v,\z) \}(s)  +s (s-1) (s-2) (s-3)\partial_{\z} \mathcal{L} \{ Q(v,\z) \}(s)+12 \z  = 0 \ .
\end{align}  
The solution is  
\begin{align}
\mathcal{L} \{ Q(v,\z) \}(s)= \frac{4 \big(s(s-1) (s-2) (s-3)  +6 s \z -9 \z \big)}{3 (3-2 s)^2}+ c_1(s) e^{\frac{3 (2 s-3) \z}{s (s-1) (s-2) (s-3)}} \ .
\end{align} 
As before, we determine the function $c_1(s)$ by examining the small $\z$ behaviour:
\begin{align}
& \lim_{\z \to 0} \mathcal{L} \{ (1-e^v)^3 F(v,\z)\}(s) = \lim_{\z \to 0} \mathcal{L} \{{Q(v,\z) \over  \z^2}\}(s) \\
&=\frac{1}{ \z^2} \Big(\frac{4 s (s-1) (s-2) (s-3)}{3 (3-2 s)^2}+c_1(s)\Big)+\frac{1}{ \z} \Big(\frac{4}{2 s-3}+\frac{3 (2 s-3) c_1(s)}{s (s-1) (s-2) (s-3)}\Big)+ {\cal O} (\z^0)  \ . \nn 
\end{align}  
To remove the divergences, set
\begin{align}
c_1(s)= -\frac{4 }{3 (3-2 s)^2} s(s-1) (s-2) (s-3) \ . 
\end{align}  
We obtain 
\begin{align}
 F(v,\z)&=   {1\over (1-e^v)^3 \z^2}  Q(v,\z)\nn\\
&= {4\over 3  (1-e^v)^3 \z^2} \mathcal{L}^{-1}\Big\{ {1\over (3-2 s)^2}\Big(s (s-1) (s-2) (s-3)+   3( 2s -3) \z \nn\\
&~~~~~~~~~~~~~~~~~~~~~~~~~~~~~~~ - s (s-1) (s-2) (s-3)  e^{\frac{3 (2 s-3) \z }{s (s-1) (s-2) (s-3)}} \Big) \Big\} (v) \ .
\end{align}   
After shifting $s \to s + {3\over 2}$, this agrees with \eqref{4Dexact2} obtained via a summation.   

\subsubsection*{$\Delta= -3$ case}

The equation \eqref{4dQeq3} can be written as 
\begin{align}
\Big(\partial^5_v \partial_{\z}-10 \partial^4_v \partial_{\z}+35 \partial^3_v \partial_{\z}-50 \partial^2_v \partial_{\z}-21 \partial^2_v + 24 \partial_v \partial_{\z} +84 \partial_v -72 \Big)Q(v,\z) = 0
\end{align} 
with constant coefficients in this form.  
 The superscript is dropped. 
After transforming the equation to the $s$-plane and imposing $Q^{(4,1)}(0,\z)= 72 \z^2$ and $Q^{(3,1)}(0,\z)=Q^{(2,1)}(0,\z)=Q^{(1,1)}(0,\z)=Q^{(1,0)}(0,\z)=Q^{(0,1)}(0,\z)=Q(0,\z)=0$, which generalise \eqref{BC}, the equation reduces to
\begin{align}
&3 \big(7 s (s -4 )+24\big)  \mathcal{L} \{ Q(v,\z) \}(s)\nn\\
&~~~~~~~~~~~~~~~~~~~~~~ - s (s-1) (s-2) (s-3) (s-4)  \partial_{\z} \mathcal{L} \{ Q(v,\z) \}(s)+ 72 \z^2 = 0 \ .
\end{align}  
The rest of the computations are essentially identical to the previous cases. Let us state the result: 
\begin{align}
 F(v,\z)&=   {1\over (1-e^v)^4 \z^3}  Q(v,\z)\nn\\
&= -{8\over 3  (1-e^v)^4 \z^3} \mathcal{L}^{-1}\Big\{ \frac{1}{\big(7 (s-4) s+24\big)^3} \Big[ 2 \kappa^2(s)   \nn\\
&~~~ +6  \kappa(s) \Big(7 (s-4) s+24\Big)  \z +9 \Big(7 (s-4) s+24\Big)^2 \z^2 - 2 \kappa^2(s)  e^{\frac{3 \left(7 (s-4 )s+24\right) \z}{\kappa(s)}}\Big] \Big\} (v) 
\end{align} 
where $\kappa(s)= s (s-1)  (s-2)  (s-3) (s-4)$.  After shifting $s\to s+2$, this result agrees with \eqref{4Dexact3}, which was obtained via a summation.  

As previously discussed, in all $\Delta=-1,-2,-3$ cases, the information about minimal-twist multi-stress tensors at large central charge is fully encoded in the term involving the essential singularities. 

\section{Correlators from modes}
\label{sec5}

In this section, we discuss an alternative approach to probing the correlator structure. By focusing on minimal-twist multi-stress tensor exchanges at large central charge, we demonstrate that the relevant part of the conformal correlators may be reconstructed from certain modes. Our derivations are based on the linear differential equations. 

\subsection{$d=2$}

Let us start with the $d=2$ case. The basic idea is to replace the $(1-s)^{-2}$ factor in the recursion relation \eqref{int2d} by a summation 
\begin{align}
\frac{1}{(1-s)^2} =\sum _{m=2}^{\infty } (m-1) s^{m-2} \ .
\end{align} 
We then write 
\begin{align}
R^{{\Delta=-1}}_{n+1}(z) = - 6  \sum _{m=2}^{\infty }  (m-1)   \int^z_0 ds_2 \int^{s_2}_0 ds_1  ~  s_1^{m-2} R^{{\Delta=-1}}_n(s_1) 
\end{align}    
with $R^{{\Delta=-1}}_0(z)= z$. 
After performing the integrals, the correlator involving the level-2 degenerate scalars with $\Delta= -1$ can be represented purely as summing over modes.

In the first few orders, we have  
\allowdisplaybreaks
\begin{align}
R^{{\Delta=-1}}_1(z)&=   \sum_{m=2}^{\infty}\frac{- 6 (m-1) z^{m+1}}{m (m+1)}\\
R^{{\Delta=-1}}_2(z)&=  \sum_{m,n=2}^{\infty}\frac{36 (m-1) (n-1) z^{m+n+1}}{m (m+1) (m+n) (m+n+1)}\\
\label{R3mode}
R^{{\Delta=-1}}_3(z)&=  \sum_{m,n,q=2}^{\infty} \frac{- 216 (m-1) (n-1) (q-1) z^{m+n+q+1}}{m (m+1) (m+n) (m+n+1) (m+n+q) (m+n+q+1)} \ .
\end{align} 
In these expressions, we use $m,n,q$ to represent modes. Higher-order expressions can be readily worked out. Recall $F(z)=z^{-1} Q(z)= z^{-1} \sum_{j}^{\infty} \eta^j R_j(z) $ in this $d=2$ case.

To evaluate these summations, we introduce the following auxiliary-variable method.  
Taking $R_{3}(z)$ as an example, we can write \eqref{R3mode} as
\allowdisplaybreaks
\begin{align}
R^{{\Delta=-1}}_3(z)&= - 216 \lim_{t_3\to 1} \int_0^{z} d{s_2}  \int_0^{s_2}  d{s_1} \int_0^{t_3} d{t_2} \int_0^{t_2} d{t_1} \nn\\
&~~~~~~~~~~~~~~~~~~\times \sum_{m,n,q=2}^{\infty} \frac{(m-1) (n-1) (q-1)}{m (m+1)}  t_1^{m+n-1} s_1^{m+n+q-1}
\end{align} 
where $t_1$ is an auxiliary variable whose role is to generate the $(m+n)(m+n+1)$ factor in the denominator via the double integrals.  
The summations over modes $m,n,q$ in this form are simpler to carry out.   After performing the  integrals, the result reproduces \eqref{R32d}.  
More auxiliary variables are required to evaluate higher-order mode summations. This method can be used to compute the similar summations in $d=4$ which we will consider later. 

After working out higher-order solutions, we identify a pattern – the $d=2$ correlator can be represented as 
\begin{align}
F^{{\Delta=-1}}(z)= 1+ \sum_{n=1}^{\infty} \sum_{\{k_i \}=2}^{\infty}  \Bigg[ \frac{ (-6)^n \prod _{i=1}^n (k_i-1)}{\prod _{l=1}^n \left(\sum _{i=1}^l k_i\right) \left(\sum _{i=1}^l k_i+1\right)} \eta^n z^{\sum _{i=1}^n k_i} \Bigg]
\end{align} 
with independent modes $k_i$ ($e.g.$, $k_1=m$).\footnote{In this expression, $n$ is the order of the expansion instead of modes.} 
As this result is derived from the differential equation, we are restricted to the level-2 degenerate field.   

We note that, in $d=2$, a set of diagrammatic rules was developed in \cite{Fitzpatrick:2015foa}, enabling the computation of the correlator via mode summations for any $\Delta$; see also Appendix D in  \cite{Fitzpatrick:2015qma} for a related computation based on the Virasoro algebra. Summing the diagrams is the same as counting the numbers of certain on-shell tree diagrams with vertices, $i.e.$, the Catalan numbers. 
The mode summations we derived from the linear differential equation are consistent with the $d=2$ diagrammatic rules. 

\subsection{$d=4$}

First we consider the simplest case with $\Delta=-1$.  Using the identity
\begin{align}
\label{mode4did}
\frac{1}{(1-s)^3} = \frac{1}{2} \sum _{m=3}^{\infty } (m-1) (m-2) s^{m-3} 
\end{align} 
we write the recursion relation \eqref{Recur4d1} as
\begin{align}
 R^{\Delta=-1}_{n+1}(z) &=  {-1 \over 2}    \sum _{m=3}^{\infty }  {(m-1)(m-2)  \over (n+2)}  \int^z_0 ds_3 \int^{s_3}_0 ds_2 \int^{s_2}_0 ds_1  ~ {s_1}^{m-3} R^{\Delta=-1}_n(s_1)
\end{align} 
with $R^{\Delta=-1}_0(z)= z^2$. 
After carrying out the integrals, we can organise the correlator structure order by order as mode summations.  
Explicit expressions in the first few orders are collected in Appendix \ref{AppB}. 
 After working out higher-order terms, a pattern emerges and we find that the corresponding correlator can be written as
{\small
\allowdisplaybreaks
\begin{align}
F^{{\Delta=-1}}(z,\z)= 1+ \sum_{n=1}^{\infty} \sum_{\{k_i \}=3}^{\infty} \Bigg[ \frac{(-1)^n}{2^n \Gamma (n+2) } \frac{\prod _{i=1}^n \Big( (k_i-1) (k_i-2)\Big) }{\prod_{l=1}^n  \Big[ \left(\sum_{i=1}^l k_i\right) \left(\sum _{i=1}^l k_i+1\right) \left(\sum_{i=1}^l k_i+2\right)  \Big] } \z^n z^{\sum _{i=1}^n k_i}   \Bigg] 
\end{align}}with independent modes $k_i$.   These $d=4$ modes exhibit a striking similarity to those in the $d=2$ case.     

The $d=4$ recursion relations for the $\Delta= -2$ and $\Delta=-3$ cases are given in \eqref{Recur4d2} and $\eqref{Recur4d3}$, respectively.  
The computations are more complicated than the $\Delta= -1$ case but the basic idea is the same – we  express the factors $(1-s)^{-2}, (1-s)^{-3}, (1-s)^{-4}$, etc, as mode summations, $e.g.$, \eqref{mode4did}.  After performing the integrals, order by order we search for a pattern. 
We find  
{\small
\allowdisplaybreaks
\begin{align}
&F^{{\Delta= -2}}(z,\z)= 1+  \sum_{n=1}^{\infty} \sum_{\{k_i \}=3}^{\infty}\Bigg[  \frac{(-3)^n }{2^{n-1}\Gamma (n+3)} \\
&~~~~~~~~~~~~~~~~ \times \frac{ \prod _{i=1}^n\Big( (k_i-1) (k_i-2) \Big) \prod _{l=1}^n \Big[\sum _{i=1}^l k_i+\sum _{i=1}^{l-1} k_i+3\Big] }{ \prod _{l=1}^n \Big[ \left(\sum _{i=1}^l k_i\right) \left(\sum _{i=1}^l k_i+1\right) \left(\sum _{i=1}^l k_i+2\right) \left(\sum _{i=1}^l k_i+3\right) \Big] }  \z^n z^{\sum _{i=1}^n k_i} \Bigg]\nn\\
\nn\\
&F^{{\Delta= -3}}(z,\z)= 1+  \sum_{n=1}^{\infty} \sum_{\{k_i \}=3}^{\infty}\Bigg[ 
\frac{-(-3)^{n+1} }{2^{n-1}\Gamma (n+4)}\\
&\times \frac{ \Big(\prod _{i=1}^n (k_i-1) (k_i-2) \Big)  \prod _{l=1}^n \Big[7 k_l \left(\sum _{i=1}^{l-1} k_i+2\right)+7 \big(\sum _{i=1}^{l-1} k_i \big)\left(\sum _{i=1}^{l-1} k_i+4\right)+2 k_l^2+24\Big]
}{\prod_{l=1}^n \Big[ \left(\sum _{i=1}^l k_i\right) \left(\sum _{i=1}^l k_i+1\right) \left(\sum _{i=1}^l k_i+2\right) \left(\sum _{i=1}^l k_i+3\right) \left(\sum _{i=1}^l k_i+4\right) \Big]  } \z^n z^{\sum_{i=1}^n k_i} \Bigg]\ .\nn
\end{align}}In Appendix \ref{AppB}, we present explicit expressions in the first few orders.

Our motivation behind these observations pertaining to the mode structures is not to evaluate the  correlator at special $\Delta$. Indeed, we already obtained the corresponding correlator in a more efficient way in the preceding section. Instead, the motivation here is that these modes may provide a path to search for the general rules governing the $d=4$ correlator more broadly. It would be interesting to develop an effective field theory to organise these higher-dimensional mode structures.

\section{Discussion}
\label{sec6}

In this note, we discuss the exact solutions to the $d=4$ linear partial differential equations proposed in \cite{Huang:2023ikg}. Assuming the validity of the equations, the exact solutions represent the resummation of all minimal-twist multi-stress tensor contributions to the heavy-light correlator, evaluated at specific values of $\Delta$, in $d = 4$ CFTs with a simple gravity dual description.  These correlator   results correspond to a holographic computation with a spherical black hole.  We find that the resummed $d=4$ correlator at large central charge has essential singularities in the $s$-plane. We also show that the correlator can be expressed as mode summations.  

There are two immediate questions: 

\noindent 1: What are the origin and the scope of validity of the $d=4$ partial differential equations? 

\noindent 2: Can one obtain the corresponding $d=4$ exact correlator at more general values of $\Delta$ using a field-theoretic method?

Let us conclude with some thoughts on how to approach these questions.

To approach the first question, it may be useful to investigate whether these differential equations can be embedded within a larger set of differential equations that incorporate the contributions from higher-twist multi-stress tensors as well.  Alternatively, it could be useful to explore whether these differential equations have an interpretation as a limit of a bulk equation of motion via holography. 

It would be interesting to construct a relevant null state in $d=4$ at large $C_T$ and, in a similar spirit to the $d=2$ BPZ approach, derive the $d=4$ null-state differential equations  from CFT first principles. To construct a relevant null state, one may explore possible higher-derivative operators in $d=4$. It is,  however, crucial to incorporate the enhancement from the heavy operators.  
Note that the $d=4$ differential equations depend on the central charge through $\mu$. 
Although this parameter can be absorbed into $\z$ in the near-lightcone limit, this does not imply that we  consider a vanishing $\mu$. It might be necessary to construct a $d=4$ analog of the $L_{-2}$ operator, which is used to construct the $d=2$ level-two null state.
It seems natural to try to construct such an object by integrating over the $d=4$ stress tensor near a two-dimensional plane with a suitable regularisation.  
However, like the $d=2$ case, this object may not have a transparent geometric interpretation. (For example, the work \cite{Huang:2021hye} attempted to isolate relevant stress-tensor OPE contributions in the near-lightcone regime. An algebraic-like structure was obtained by assuming a certain limiting procedure. It might be interesting to see if such a structure has an observable consequence, or to consider possible deformations.)

To address the second question, it may be useful to investigate whether the $d=4$ correlator evaluated at $\Delta=-1, -2, -3$ exhibits a pattern that holds for a broader range of $\Delta$. 
In $d=2$, an analytic continuation was employed in \cite{Fitzpatrick:2015foa} to compute the  correlator at general $\Delta$ using the correlator involving degenerate fields. 
It would be interesting to explore whether a similar approach could be extended to $d=4$.  
On the other hand, it is desirable to establish a connection between our work and the effective field theory approaches developed in \cite{Cotler:2018zff, Haehl:2019eae}. To compute the multi-stress tensor contributions in higher-dimensional CFTs, developing a generalised effective field theory is of great interest.

 It is also important to emphasise that our results do not account for the double-trace contributions made from two light scalars. It would be interesting to obtain the corresponding exact correlators in higher dimensions. 

These investigations are left to future work.

\vspace{-0.3cm}

\subsection*{Acknowledgements}

I would like to thank John Cardy, Felix Haehl, Shota Komatsu, Kostas Skenderis,  Petar Tadi\'c and Balt van Rees for related discussions and comments. This research was supported in part by the UKRI grant EP/X030334/1. 

\newpage

\begin{appendices}

\section{} 
\label{AppA}

In this appendix, we collect some ${\cal O}(n)$ expressions in both $d=2$ and $d=4$.

\vspace{-0.2cm}

\subsection*{$\bullet$ $d=2$ correlator at $\Delta=-1$}

\vspace{-0.4cm}

{\small
\begin{align}
& R^{{\Delta=-1}}_0(v)= 1-e^v \nn\\
& R^{{\Delta=-1}}_1(v)= 6 \Big(v+2+e^v (v-2)\Big) \nn\\
& R^{{\Delta=-1}}_2(v)=18 \Big(v^2+6 v+12-e^v \left(v^2-6 v+12\right)\Big)\nn\\
& R^{{\Delta=-1}}_3(v)=36 \Big(v^3+12 v^2+60 v+120+e^v \left(v^3-12 v^2+60 v-120\right)\Big)\nn\\
& R^{{\Delta=-1}}_4(v) = 54 \Big(v^4+20 v^3+180 v^2+840 v+1680 -e^v \left(v^4-20 v^3+180 v^2-840 v+1680\right)\Big)\nn\\
& R^{{\Delta=-1}}_5(v) = \frac{324}{5} \Big(v^5+30 v^4+420 v^3+3360 v^2+15120 v+30240\nn\\
&~~~~~~~~~~~~~~~~~ + e^v \left(v^5-30 v^4+420 v^3-3360 v^2+15120 v-30240\right)\Big)\nn\\
& R^{{\Delta=-1}}_6(v) = \frac{324}{5} \Big( v^6+42 v^5+840 v^4+10080 v^3+75600 v^2+332640 v+665280\nn\\
& ~~~~~~~~~~~~~~~~~ -e^v \left(v^6-42 v^5+840 v^4-10080 v^3+75600 v^2-332640 v+665280\right)  \Big)\nn\\
& R^{{\Delta=-1}}_7(v) = \frac{1944}{35} \Big(v^7+56 v^6+1512 v^5+25200 v^4+277200 v^3+1995840 v^2+8648640 v+17297280\nn\\
&~~~~~~~~~~~~~~~~~~ + e^v \left(v^7-56 v^6+1512 v^5-25200 v^4+277200 v^3-1995840 v^2+8648640 v-17297280\right) \Big)  \nn 
\end{align}}

\subsection*{$\bullet$ $d=4$ correlator at $\Delta=-1$}

\vspace{-0.4cm}

{\small
\allowdisplaybreaks
\begin{align}
& R^{{\Delta=-1}}_0(v)= (1-e^v)^2 \nn\\
& R^{{\Delta=-1}}_1(v)= \frac{1}{4}  \Big(v+3+e^{2 v} (v-3)+4 e^v v\Big)\nn\\
& R^{{\Delta=-1}}_2(v)= \frac{1}{48} \Big(v^2+9 v+24-8 e^v \big(v^2+6\big)+e^{2 v} \big(v^2-9 v+24\big)\Big)\nn\\
& R^{{\Delta=-1}}_3(v)= \frac{1}{1152} \Big(v^3+18 v^2+123 v+315+16 e^v \big(v^2+24\big) v +e^{2 v} \big(v^3-18 v^2+123 v-315\big)\Big)\nn\\
& R^{{\Delta=-1}}_4(v) = \frac{1}{46080} \Big(v^4+30 v^3+375 v^2+2295 v+5760 -32 e^v \big(v^4+60 v^2+360\big)\nn\\
&~~~~~~~~~~~~~~~~~~~~~~~~~~~~ +e^{2 v} \big(v^4-30 v^3+375 v^2-2295 v+5760\big)\Big)\nn\\
& R^{{\Delta=-1}}_5(v) =  \frac{1}{2764800}  \Big(v^5+45 v^4+885 v^3+9450 v^2+54495 v+135135 \nn\\
&~~~~~~~~~~~~ +64 e^v \big(v^4+120 v^2+2520\big) v +e^{2 v} \big(v^5-45 v^4+885 v^3-9450 v^2+54495 v-135135\big)\Big)\nn\\
& R^{{\Delta=-1}}_6(v) = \frac{1}{232243200} \Big(v^6+63 v^5+1785 v^4+28980 v^3+283185 v^2\nn\\
& ~~~~~~~~~~~~~~~ +1573425 v+3870720 -128 e^v \big(v^6+210 v^4+10080 v^2+60480\big) \nn\\
&~~~~~~~~~~~~~~~ +e^{2 v} \big(v^6-63 v^5+1785 v^4-28980 v^3+283185 v^2-1573425 v+3870720\big)\Big) \nn\\
& R^{{\Delta=-1}}_7(v) = \frac{1}{26011238400} \Big(v^7+84 v^6+3234 v^5+73710 v^4+1070685 v^3+9882810 v^2\nn\\ 
& ~~~~~~~~~~~~~~~~~ +53531415 v+130945815+256 e^v \big(v^6+336 v^4+30240 v^2+604800\big) v +e^{2 v} \big(v^7 \nn\\
&~~~~~~~~~~~~~~~~~ -84 v^6 +3234 v^5-73710 v^4+1070685 v^3-9882810 v^2+53531415 v-130945815\big)\Big)\nn 
\end{align}}

\subsection*{$\bullet$ $d=4$ correlator at $\Delta=-2$}

\vspace{-0.4cm}

{\small
\allowdisplaybreaks
\begin{align}
& R^{{\Delta=-2}}_0(v)= (1-e^v)^3 \nn\\
& R^{{\Delta=-2}}_1(v)= \frac{1}{2} \big(1-e^v\big) \big(v+3+4 e^v v+e^{2 v} (v-3)\big)  \nn\\
& R^{{\Delta=-2}}_2(v)= \frac{1}{96} \Big(9 v^2+75 v+185 -27 e^v \big(v^2-5 v+5\big)+27 e^{2 v} \big(v^2+5 v+5\big) \nn\\
&~~~~~~~~~~~~~~~~~~~~~~ +e^{3 v} \big(-9 v^2+75 v-185\big)\Big)  \nn\\
& R^{{\Delta=-2}}_3(v)= \frac{1}{320} \Big(3 v^3+48 v^2+293 v+675+9 e^v\big(v^3-12 v^2+39 v-75\big) \nn\\
&~~~~~~~~~~~~~~~~~~~ -9 e^{2 v}\big(v^3+12 v^2+39 v+75\big)+e^{3 v} \big(-3 v^3+48 v^2-293 v+675\big)\Big)  \nn\\
& R^{{\Delta=-2}}_4(v) = \frac{1}{15360} \Big(9 v^4+234 v^3+2553 v^2+13749 v+30631-27 e^v \big(v^4-22 v^3+153 v^2\nn\\ 
&~~~~~~~~~~~~~~~~~~~~~~~~~~~ -603 v+1179\big) +27 e^{2 v} \big(v^4+22 v^3+153 v^2+603 v+1179\big)\nn\\
& ~~~~~~~~~~~~~~~~~~~~~~~~~~~ +e^{3 v} \big(-9 v^4+234 v^3-2553 v^2+13749 v-30631\big)\Big) \nn\\
& R^{{\Delta=-2}}_5(v) = \frac{1}{358400}  \Big(9 v^5+345 v^4+5825 v^3+53850 v^2+271205 v+592515 \nn\\
&~~~~~~~~~~~~~~~~~~~~~~~~~~~~~ +27 e^v \big(v^5-35 v^4+425 v^3-2850 v^2+12045 v-21945\big)\nn\\
&~~~~~~~~~~~~~~~~~~~~~~~~~~~~~ -27 e^{2 v} \big(v^5+35 v^4+425 v^3+2850 v^2+12045 v+21945\big) \nn\\
&~~~~~~~~~~~~~~~~~~~~~~~~~~~~~ +e^{3 v} \big(-9 v^5+345 v^4-5825 v^3+53850 v^2-271205 v+592515\big)\Big) \nn\\
& R^{{\Delta=-2}}_6(v) = \frac{1}{103219200} \Big(81 v^6+4293 v^5+103140 v^4+1431720 v^3+12063645 v^2+58288035 v \nn\\
&~~~~~~~~~~~~~~~~~~ +125737265 -243 e^v \big(v^6-51 v^5+960 v^4-9960 v^3+68445 v^2-279045 v+514485\big)\nn\\
&~~~~~~~~~~~~~~~~~~ +243 e^{2 v} \big(v^6+51 v^5+960 v^4+9960 v^3+68445 v^2+279045 v+514485\big)+e^{3 v} \big(-81 v^6\nn\\
&~~~~~~~~~~~~~~~~~~ +4293 v^5-103140 v^4+1431720 v^3-12063645 v^2+58288035 v-125737265\big)\Big) \nn\\
& R^{{\Delta=-2}}_7(v) = \frac{1}{4335206400} \Big(81 v^7+5670 v^6+183330 v^5+3537450 v^4+43855245 v^3+348298860 v^2\nn\\
&~~~~~~~~~~~~~~~~~~ +1636264175 v+3498489225 
+243 e^v \big(v^7-70 v^6+1890 v^5-28350 v^4+284445 v^3\nn\\
&~~~~~~~~~~~~~~~~~~ -1905120 v^2 +7649775 v-14397075\big)
-243 e^{2 v} \big(v^7+70 v^6+1890 v^5+28350 v^4 \nn\\
&~~~~~~~~~~~~~~~~~~ +284445 v^3 +1905120 v^2 +7649775 v+14397075\big)+e^{3 v} \big(-81 v^7+5670 v^6  \nn\\
&~~~~~~~~~~~~~~~~~~ -183330 v^5 +3537450 v^4 -43855245 v^3+348298860 v^2-1636264175 v+3498489225\big)\Big) \nn 
 \end{align}}

\subsection*{$\bullet$ $d=4$ correlator at $\Delta=-3$}

\vspace{-0.4cm}

{\small
\allowdisplaybreaks
\begin{align}
&  R^{{\Delta=-3}}_0(v)= (1-e^v)^4 \nn\\
&  R^{{\Delta=-3}}_1(v)= \frac{3}{4} \left(1-e^v\right)^2 \big(v+3 +4 e^v v+e^{2 v} (v-3)\big)   \nn\\
& R^{{\Delta=-3}}_2(v)=  \frac{1}{80} \Big(18 v^2+141 v+325-2 e^v \left(9 v^2-123 v+176\right)\nn\\
&~~~~~~~~~~~~~~~~~~~~~ +54 e^{2 v} \left(2 v^2+1\right)-2 e^{3 v} \left(9 v^2+123 v+176\right)+e^{4 v} \left(18 v^2-141 v+325\right)\Big)\nn\\
& \label{R3} R^{{\Delta=-3}}_3(v)= \frac{1}{160} \Big(6 v^3+87 v^2+481 v+1006 +e^v \left(3 v^3-96 v^2+647 v-857\right) -18 e^{2 v} v \left(2 v^2-3\right)\nn\\
&~~~~~~~~~~~~~~~~~~~~~~~ +e^{3 v} \left(3 v^3+96 v^2+647 v+857\right)+e^{4 v} \left(6 v^3-87 v^2+481 v-1006\right)\Big) \nn\\
&  R^{{\Delta=-3}}_4(v) = \frac{1}{17920}\Big( 72 v^4+1656 v^3+16014 v^2+76743 v+152972 -2 e^v \big(9 v^4-522 v^3+8295 v^2\nn\\
&~~~~~~~~~~~~~~~~~-40857 v+68224\big)+216 e^{2 v} \left(2 v^4-18 v^2-153\right)-2 e^{3 v} \big(9 v^4+522 v^3+8295 v^2\nn\\
&~~~~~~~~~~~~~~~~~+40857 v+68224\big) +e^{4 v} \left(72 v^4-1656 v^3+16014 v^2-76743 v+152972\right)\Big)\nn\\
&  R^{{\Delta=-3}}_5(v) = \frac{3}{1433600} \Big(144 v^5+4800 v^4+70700 v^3+572850 v^2 +2543330 v+4930765  \nn\\
&~~~~~~~~~~~~~~~~~ +2 e^v \left(9 v^5-825 v^4+23825 v^3-263100 v^2+1185305 v-2378215\right) -432 e^{2 v} v \big(2 v^4  \nn\\
&~~~~~~~~~~~~~~~~~  -50 v^2-885\big) +2 e^{3 v} \left(9 v^5+825 v^4+23825 v^3+263100 v^2+1185305 v+2378215\right)\nn\\
&~~~~~~~~~~~~~~~~~ +e^{4 v} \left(144 v^5-4800 v^4+70700 v^3-572850 v^2+2543330 v-4930765\right)\Big)\nn\\
&  R^{{\Delta=-3}}_6(v) = \frac{1}{154828800} \Big(2592 v^6+117936 v^5+2441880 v^4+29358000 v^3+215505990 v^2+913040835 v \nn\\
&~~~~~~~~~~~~~~~+1738983400 -2 e^v \big(81 v^6-10773 v^5+492345 v^4-9702000 v^3+88522245 v^2-404456955 v \nn\\
&~~~~~~~~~~~~~~~+870410240\big) +3888 e^{2 v} \left(4 v^6-210 v^4-5670 v^2+945\right)-2 e^{3 v} \big(81 v^6+10773 v^5+492345 v^4\nn\\
&~~~~~~~~~~~~~~~+9702000 v^3+88522245 v^2+404456955 v+870410240\big) +e^{4 v} \big(2592 v^6-117936 v^5\nn\\
&~~~~~~~~~~~~~~~+2441880 v^4-29358000 v^3+215505990 v^2-913040835 v+1738983400\big)\Big) \nn\\
& R^{{\Delta=-3}}_7(v) 
= \frac{1}{3612672000} \Big(2592 v^7+154224 v^6+4256280 v^5+70454160 v^4+753629310 v^3\nn\\
&~~~~~~~~~~~~~~~ +5196824115 v^2+21338803990 v+40150046895  +e^v \big(81 v^7-14742 v^6+978264 v^5\nn\\
&~~~~~~~~~~~~~~~-30248190 v^4+472613085 v^3 -3953470920 v^2+18595335185 v-40478016495\big)\nn\\
&~~~~~~~~~~~~~~~-3888 e^{2 v} v \big(4 v^6-378 v^4-13230 v^2+57645\big) +e^{3 v} \big(81 v^7+14742 v^6 +978264 v^5\nn\\
&~~~~~~~~~~~~~~~+30248190 v^4+472613085 v^3+3953470920 v^2+18595335185 v+40478016495\big)\nn\\
&~~~~~~~~~~~~~~~+e^{4 v} \big(2592 v^7-154224 v^6+4256280 v^5-70454160 v^4+753629310 v^3\nn\\
&~~~~~~~~~~~~~~~-5196824115 v^2+21338803990 v-40150046895\big)\Big)\nn
\end{align}} 

\section{} 
\label{AppB}

This appendix includes some explicit expressions of the $d=4$ modes at $\Delta=-1, -2, -3$:
{\footnotesize
\allowdisplaybreaks
\begin{align}
F^{{\Delta= -1}}(z,\z)&=1 
- \sum_{m=3}^{\infty} \frac{(m-1) (m-2) z^{m} \z }{4 m (m+1) (m+2)} \nn\\
&~~~~~ + \sum_{m,n=3}^{\infty}\frac{ (m-1) (m-2) (n-1) (n-2)  z^{m+n} \z^2}{24 m (m+1) (m+2) (m+n) (m+n+1) (m+n+2)}  \nn\\
& ~~~~~ -\sum_{m,n,q=3}^{\infty} \Bigg( \frac{(m-1) (m-2) (n-1) (n-2)  }{192 m (m+1) (m+2) (m+n) (m+n+1) (m+n+2)}\nn\\
&~~~~~~~~~~~~~~~~~~~~~ \times \frac{ (q-1) (q-2)  z^{m+n+q}\z^3  }{ (m+n+q) (m+n+q+1) (m+n+q+2)} \Bigg)  + {\cal O} (\z^4) \nn\\
F^{{\Delta= -2}}(z,\z)&=1 
-\sum_{m=3}^{\infty} \frac{(m-1) (m-2) z^m \z }{2 m (m+1) (m+2)}\nn\\
&~~~~~~ +\sum_{m,n=3}^{\infty} \frac{3 (m-1) (m-2) (n-1) (n-2) (2 m+n+3) z^{m+n} \z^2 }{16 m (m+1) (m+2) (m+n) (m+n+1) (m+n+2) (m+n+3)}\nn\\
&~~~~~~ -\sum_{m,n,q=3}^{\infty} \Bigg(\frac{9 (m-1) (m-2) (n-1) (n-2)  (2m+n+3) }{160 m (m+1) (m+2) (m+n) (m+n+1) (m+n+2) (m+n+3) }\nn\\
&~~~~~~~~~~~~~~~~~~~~~ \times\frac{ (q-1) (q-2) (2 m+2 n+q+3) z^{m+n+q} \z^3  }{ (m+n+q) (m+n+q+1) (m+n+q+2) (m+n+q+3)}\Bigg) + {\cal O} (\z^4) \nn\\
F^{{\Delta= -3}}(z,\z)&= 1 -\sum_{m=3}^{\infty}  \frac{3 (m-1) (m-2)  z^m \z}{4 m (m+1) (m+2)}\nn\\
&~~~~~~ + \sum_{m,n=3}^{\infty}  \frac{9 (m-1) (m-2) (n-1) (n-2)  \big(7 m^2+7 m (n+4)+2 \left(n^2+7 n+12\right)\big) z^{m+n} \z^2}{40 m (m+1) (m+2) (m+n) (m+n+1) (m+n+2) (m+n+3) (m+n+4)}\nn\\
&~~~~~~ -\sum_{m,n,q=3}^{\infty}\Bigg( \frac{9 (m-1) (m-2) (n-1) (n-2)  \left(7 m^2+7 m (n+4)+2 \left(n^2+7 n+12\right)\right)  }{160 m (m+1) (m+2) (m+n) (m+n+1) (m+n+2) (m+n+3) (m+n+4) }\nn\\
&~~~~~~~~~
\times \frac{ (q-1) (q-2)  \big(7 m^2+7 m (2 n+q+4)+7 n^2+7 n (q+4)+2 (q^2+7 q+12)\big) z^{m+n+q}  \z^3}{(m+n+q) (m+n+q+1) (m+n+q+2) (m+n+q+3) (m+n+q+4)}\Bigg)  + {\cal O} (\z^4)\nn
\end{align}}

\end{appendices}

\bibliographystyle{utphys}  
\bibliography{resumTs}  

\end{document}